\documentclass[a4paper,11pt]{article}

\pdfoutput=1 
\usepackage{jcappub} 

\usepackage[T1]{fontenc} 

\usepackage{amssymb}
\usepackage[dvipsnames]{xcolor}
\usepackage{multirow}
\usepackage{bm}
\usepackage{verbatim}
\usepackage{amsmath}
\usepackage{siunitx}
\usepackage{hyperref}

\newcommand{\be}{\begin{equation}}
\newcommand{\ee}{\end{equation}}
\newcommand{\nn}{\nonumber}
\newcommand{\bea}{\begin{eqnarray}}
\newcommand{\eea}{\end{eqnarray}}
\newcommand{\M}{\mathcal{M}}
\renewcommand{\d}{{\rm d}}
\newcommand{\dd}{\partial}

\newcommand{\bn}{{\bm n}}

\newcommand{\de}{\delta}
\newcommand{\De}{\Delta}

\everymath{\displaystyle}

\newcommand{\refrep}[1]{\textcolor{black}{#1}}

\title{\boldmath Magnification and evolution bias of transient sources: GWs and SNIa}

\author[a,1]{Stefano Zazzera,\note{Corresponding author.}}
\author[a,b,c,d]{José Fonseca,}
\author[a,e]{Tessa Baker}
\author[a,d,f]{and Chris Clarkson}

\affiliation[a]{Department of Physics \& Astronomy, Queen Mary University of London, Mile End Road, London E1 4NS, United Kingdom}
\affiliation[b]{Instituto de Astrofisica e Ci\^{e}ncias do Espa\c{c}o, Universidade do Porto CAUP, Rua das Estrelas, PT4150-762 Porto, Portugal}
\affiliation[c]{Departamento de F\'isica e Astronomia, Faculdade de Ci\^{e}ncias, Universidade do Porto, Rua do Campo Alegre 687, PT4169-007 Porto, Portugal}
\affiliation[d]{Department of Physics \& Astronomy, University of the Western Cape, Cape Town 7535, South Africa}
\affiliation[e]{Institute of Cosmology and Gravitation, University of Portsmouth, Burnaby Road, Portsmouth PO1 3FX, UK}
\affiliation[f]{Department of Mathematics \& Applied Mathematics, University of Cape Town, Cape Town 7701, South Africa}

\emailAdd{s.zazzera@qmul.ac.uk}
\emailAdd{jose.fonseca@astro.up.pt}
\emailAdd{tessa.baker@port.ac.uk}
\emailAdd{chris.clarkson@qmul.ac.uk}

\abstract{Third-generation gravitational wave (GW) observatories such as the Einstein Telescope and Cosmic Explorer, together with the LSST survey at the Vera Rubin Observatory, will yield an abundance of extra-galactic transient objects. This opens the exciting possibility of using GW sources and Supernovae Type Ia (SNIa) as luminosity distance tracers of large-scale structure for the first time.
The large volumes accessible to these surveys imply that we may need to include relativistic corrections, such as lensing and Doppler magnification. However, the amplitude of these effects depends on the magnification and evolution biases of the transient sources, which are not yet understood.
In this paper we develop comprehensive frameworks to address and model these biases for both populations of transient objects; in particular, we define how to compute these biases for GW sources. We then analyse the impact of magnification and evolution biases on the relativistic corrections and on the angular power spectrum of these sources. We show that correct modelling and implementation of these biases is crucial for measuring the cross-correlations of transient sources at higher redshifts.}

\begin{document}
\maketitle
\flushbottom

\section{Introduction}
Clustering analysis of cosmological sources to understand the large-scale structure of the Universe has been explored extensively for both galaxy surveys and intensity mapping (IM) in redshift space \cite{Martinez_1999,Padmanabhan_2007,Bonvin_2008,Hui_2008,Chen_2008,Yoo_2009,Bonvin_2011,Challinor_2011,Jeong_2012,Tansella_2018,Castorina_2022}, as current surveys 
are dominated by these kinds of tracers. The observed density contrast
depends upon an extensive study of various effects, including Redshift Space Distortions (RSDs) \cite{Kaiser_1992,Hamilton_1998}, weak lensing \cite{Sloan}, relativistic corrections \cite{Challinor_2011,Bonvin_2011} and properties of the tracer considered, e.g. clustering bias \cite{Jeong_2012}, luminosity function and redshift distribution \cite{Jain_2011,Mason_2015,Leung_2018,Duncan_2022}. 

However, the upcoming generation of surveys and gravitational wave experiments will provide us with an unprecedented number of extra-galactic transient objects such as Supernovae Type Ia (SNIa) and gravitational waves (GWs). In particular, the Legacy Survey of Space and Time (LSST) is forecasted to access millions of photometric SNIa \cite{Libanore_2022,LSST_2021,Sanchez_2022,LSST_book}, and similar numbers of gravitational waves emitted by compact object mergers will be observed by the ground-based gravitational wave observatories Einstein Telescope (ET) \cite{Libanore_2022, Libanore_2021, Scelfo_2018, Scelfo_2020, Scelfo_2022, Scelfo_2022_2, Sathyaprakash_2010, Sathyaprakash_2012, Maggiore_2020, Punturo_2010} and Cosmic Explorer (CE) \cite{Evans_2021, Reitze_2019, Mpetha_2022}. 

Along with the sheer number of objects seen, this new generation of observatories will push the horizon of detections further: whilst LSST will observe SNIa events at $z < 4$ \cite{Sathyaprakash_2010, Sanchez_2022, LSST_book}, ET and CE are predicted to access virtually all binary black hole \refrep{(BBH)} mergers up to $z \sim 10$ \cite{Sathyaprakash_2010, Maggiore_2020, Punturo_2010, Evans_2021, Peron_2023}. Greatly increasing the source distances also increases the relevance of relativistic correction effects. This is due to both the impact of lensing on the larger redshifts covered, and due to tracers evolving with cosmic time, thus not distributing across different redshift bins equally \cite{Rigby_2011,Duncan_2022}. 

These future prospects indicate that the kinds of clustering analyses carried out with galaxies and IM will soon be applicable to transient events such as SNIa and GWs too. A key subtlety is that the former are tracers living in redshift space, whilst the latter carry distance information only under the form of a luminosity distance. 
In a previous work 
\cite{Us} we explored the difference between clustering analysis in redshift space and in luminosity distance space for a generic tracer. We found that the two can be significantly different, leading to large discrepancies in the corresponding angular power spectrum, up to $50\%$ at large scales. Therefore, any analysis utilising tracers such as GWs or SNIa should be carried out in luminosity distance.

Primarily, one is required to build an expression for the observed number counts in luminosity distance space, which not only traces the underlying density of matter, but also includes distortion effects along our past light-cone. One can write the generic expression for the density contrast in luminosity distance space as \cite{Us,Namikawa}:
\be \label{eq:general_num_contrast} 
\De_O(D_L,\hat{n}) = \de_n-\left[3-b_e-\frac{5}{\gamma}s\right]\gamma\frac{\de D_L}{D_L}+\frac{\de V(\bn,D_L)}{V(D_L)} \, \,
\ee
where $\gamma \equiv r\mathcal{H}/1+r\mathcal{H}$ with comoving distance $r$ and comoving Hubble \refrep{parameter} $\mathcal{H}$, $\delta_n$ is the density contrast in the Newtonian gauge, $D_L$ the luminosity distance, and $V$ the volume. The distortion is then found to be dependent on two extra functions, the evolution bias \cite{Maartens_2021}
\be \label{eq:be_def}
b_e \equiv \frac{\partial\ln n}{\partial\ln a}\bigg|_{d_{th}}\,,
\ee
where \refrep{$d_{th}$ is the detector's observation threshold (taken as the luminosity cut $L_c$ in the standard galaxy case and signal-to-noise ratio threshold $\rho_{th}$ for GWs)}, and the magnification bias $s$, defined as the change in the comoving number density $n$ at fixed redshift/distance with respect to the luminosity cut. Instead, $b_e$ is the change in the comoving number density with respect to the scale factor, while keeping the detector's threshold fixed \cite{Maartens_2021}. 
Physically, the magnification bias accounts for objects being magnified in or out of the detector's flux limit as a result of a perturbation, and the evolution bias describes the impact on clustering of a (possibly) non-conserved comoving number density through cosmic time. A null evolution bias would imply a constant observed number density through redshift, whilst a non-zero one corrects each redshift bin for the effect of an evolving population. In other words, it traces how the observed number of objects per unit volume changes as the universe expands.

We define the magnification bias for GWs and SNIa differently in order to preserve the general expression in eq. \eqref{eq:general_num_contrast}: the former in terms of signal-to-noise ratio (SNR) threshold and the latter in terms of magnitude cut, respectively. In the following sections we will provide the expression for each tracer, as well as specify the magnification bias for different surveys and thresholds. 



Here, we propose a theoretical framework to model
both GWs and SNIa, thus aiming at consistently translating terminology and modelling previously used in galaxy clustering into equivalent expressions applicable to these new transient tracers. One should note that some of these biases, at least for GWs, were recently explored for the first time in \cite{Scelfo_2018,Scelfo_2020,Scelfo_2022, Mastrogiovanni_2023}, particularly in the context of cross-correlations, as these have become increasingly exciting with the prospect of future detectors.

The paper is structured as follows. In section \ref{sec:gw} we describe the modelling of the biases for GWs, justifying our
theoretical formalism 
and producing the required event rate and chirp mass distributions. 
Further, we model the biases for SNIa in section \ref{sec:SNIa}. Finally, we apply these biases to the kernels of the number count fluctuation and the angular power spectrum in section \ref{sec:applications}, and study their impact. Section \ref{sec:conclusions} is then devoted to summary and conclusions.

\section{GW Biases}\label{sec:gw}

Here we illustrate the modelling of the magnification and evolution biases for GWs. We first explain the reasoning behind choosing the signal-to-noise ratio instead of luminosity as discriminant for the magnification bias; we then describe both the modelling for the event rate and the chirp mass distribution, which we derive using the primary mass distribution from \cite{LIGOScientific:2021djp,GWTC3,GWTC2}. An interested reader can find the distributions of primary and secondary masses in Appendix \ref{sec:mass_distributions}, and the derivation of the chirp mass probability distribution function in Appendix \ref{sec:chirp_calc}. We then evaluate the GW biases for the different mass models used. 

For galaxy surveys the magnification bias is generally defined from \cite{Maartens_2021} as:
\begin{equation}\label{eq:mag_bias}
    s(a,L_{c}) = -\frac{\partial \ln n_{g}(a,L_{c})}{\partial \ln L_{c}}\bigg|_a
\end{equation}
where $L_{c}$ is the luminosity threshold of our survey at each redshift and $n_{g}(a,L_{c})$ is the comoving number density of sources above the threshold. The latter is defined by integrating the (comoving) luminosity function $\Phi$ over luminosity:
\be
    n_{g}(a,L_{c}) = \int_{L_{c}(a)}^{\infty } \d L\ \frac{\Phi(a,L)}{L_{\ast}(a)}
\ee
where $L_{\ast}$ is a characteristic luminosity in the luminosity function.
The number of sources that are \textit{above} the corresponding luminosity cut is the same as the number of sources that are \textit{observed}. This implies that regardless of the choice of luminosity cut, $n_{g}$ can be zero at a certain redshift, as there will be no sources emitting at a sufficiently high luminosity. Therefore, the value for the magnification bias is expected to increase with redshift up until its validity limit, i.e. where sources are not detectable anymore, meaning that observing fewer and fewer sources leads to a (positive) larger value of $s$.

We can now transport this concept from the galaxy case to the GW scenario. However, for GWs we can't produce a detector-independent quantity analogous to EM luminosity, as the quality of a detection (the SNR, $\rho$) is inherently tied to the one-sided power spectral density of the detector, i.e. its sensitivity, $PSD(f)$ \cite{Finn_1996}:
\be
   \rho^{2} = 4 \int \d f\  \frac{|\tilde{h}(f)|^{2}}{PSD(f)}.
\ee
In galaxy surveys the samples are also telescope dependent, as different telescopes target different parts of the spectrum or optimize for different types of galaxies. As an example Euclid will target H$\alpha$ emitting galaxies, while DESI will have broader target to any emission line. The response function of a GW detector is equivalent to the instrumental design of optical and near-infrared telescopes. 
Using the characteristic strain $h(f)$ of a GW event (which depends on chirp mass $\mathcal{M}$ and redshift) is not possible either, as it still has to be compared to the detector's sensitivity curve, and their ratio integrated to determine whether the event is detected and at which level of confidence. 
The two scenarios would be similar if we assumed the sensitivity curve of GWs detector to be flat across all frequencies, and the GWs signal to be a single-frequency burst. All signals would then see the same response from the detector, as they would not move across frequency space. 

We therefore define the magnification bias for GW merger events in term of the SNR as
\be \label{eq:gw_mag}
s^{GW}\equiv -\frac1{5}\frac{\partial\ln n}{\partial\ln\rho}\bigg|_a\,.
\ee
The factor of $\frac{1}{5}$ comes from implicitly fixing the expression eq. \eqref{eq:general_num_contrast} and setting $s$ accordingly; this was done to keep eq. \eqref{eq:general_num_contrast} general, particularly for coding purposes. We therefore need to model the number density of observed events as a function of the detector signal-to-noise ratio $\rho$. Eq. \eqref{eq:gw_mag} defines the magnification bias for any GW source. However, for the purpose of this paper we will only focus on GWs from binary black hole mergers. The same method can be applied to binary neutron stars and neutron star black hole pairs, if the appropriate number density is provided.

\subsection{Event rate for GWs}\label{sec:event}
The number density of observed sources is the number above a certain SNR threshold, which is usually assumed to be $\rho_{th}=8$ \refrep{for a single detector. In the case of multiple detectors, $\rho_{th}$ is added in quadrature.}

\begin{figure}
\centering
\includegraphics[width=\textwidth]{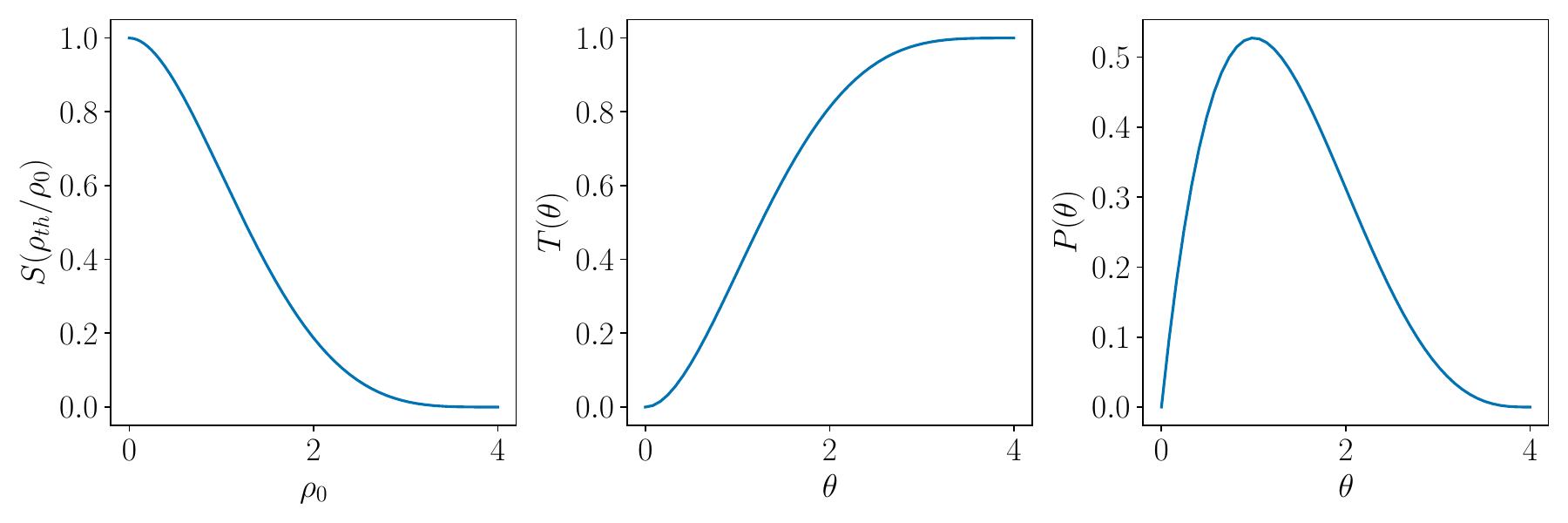}
\caption{{\em Left:} Fraction of sources detected, $S(\rho_{th}/\rho_{0})$ as a function of $\rho_0$, i.e. the characteristic SNR of the source. Low values of $S$ imply either a low SNR threshold or loud events, whereas large $S$ signifies a high threshold or quiet events. {\em Center:} Fraction of sources undetected at each random orientation. $\theta$ is equivalent to the ratio between the SNR threshold $\rho_{th}$ and the characteristic SNR of the source, $\rho_0$. {\em Right:} Orientation PDF. }
\label{fig:theta}
\end{figure}

From \cite{Oguri_2018}, we can model the number density of BBH mergers \refrep{in comoving volume} as:
\begin{equation}\label{eq:GW_numdens}
    n_{\rm GW}(z) = \refrep{\tau}\frac{R_{GW}(z)}{1+z}\int \d \mathcal{M}\ \phi(\mathcal{M})S(\rho_{th};\mathcal{M},z)
\end{equation}
\refrep{where $\tau$ is the observation time of the detector}, $R_{GW}(z)$ is the intrinsic merger rate, $\phi(\M)$ is the chirp mass $\M$ distribution, with
\bea \label{eq:chirp_def}
\M \equiv \frac{(m_1m_2)^{3/5}}{(m_1+m_2)^{1/5}} \, .\
\eea
Selection effects of the signal-to-noise threshold are included through the implementation of a survival function $S(\rho_{th};\M,z)$, which is defined as the fraction of sources that are above the SNR threshold, $\rho_{th}$, at any given mass and redshift bin \footnote{We assume a theoretical optimal SNR. While there should be a difference between the optimal value of $\rho$ and the detected one, we expect it to not qualitatively impact the results of our work, although a more accurate approach should be taken when dealing with real data}. \refrep{We note that for physical experiments, the sensitivity curve should be defined on the detected SNR, rather than the optimal one. The former has a non-central $\chi^2$ distribution \cite{referee} which depends on the optimal SNR, which by itself is a function of the PSD. Although our simplified assumption is not expected to qualitatively change the result of this study, a rigorous analysis with real data would have to take this into account for a more realistic experiment.}
To evaluate $S(\rho_{th};\M,z)$, we start from an expression for the SNR $\rho$ for a compact binary merger\cite{Oguri_2018,Finn_1996}:
\bea\label{eq:theta}
    \rho &=& \sqrt{\frac5{96\pi^{4/3}}} \frac1{D_L}[(1+z)\mathcal{M}]^{5/6}\theta\sqrt{I}=\rho_0(\mathcal{M},z)\,\theta \, ,\\
    I &=& \int_0^{f_{max}} \d f\ f^{-7/3}\left[PSD(f)\right]^{-1} \, ,\
\eea
where $\theta$ is the orientation of the binary with respect to the detector, and $\rho_0(\mathcal{M},z)$ encapsulates signal and detector's response. \refrep{For simplicity, here we only use a 0PN approximation for the signal to compute the SNR. A more rigorous analysis would include higher order corrections to describe merger and ringdown parts without this approximation, thus including a boost to the SNR.} Assuming random orientations, the PDF of $\theta$ can be well approximated by \cite{Finn_1996}:
\be \label{eq:Ptheta}
P(\theta) = \frac{5\theta}{256}(4-\theta)^{3} \, ,\
\ee
for $0 < \theta < 4$ and $P(\theta) = 0$ otherwise. \refrep{Note that this is valid for a single L-shaped detector. In the case of three detectors such as LVK, the SNR can be added in quadrature, although for a triangular configuration such as ET, this would change. However, for simplicity we assume the same function applies to an ET-like experiment.}
Further, we assume that $f_{max}$ corresponds to the frequency at the innermost stable circular orbit:
\bea \label{eq:fmax}
f_{max} \simeq \frac{4397 \text{Hz}}{(1+z)(M/M_{\odot})} \, .\
\eea
From eq. \eqref{eq:theta} we have $\theta=\rho/\rho_0$, thus by fixing $\rho=\rho_{th}$ we can obtain a PDF for the sources that produce a sufficiently high SNR to be detected. The corresponding cumulative distribution function is the integral of eq.  \eqref{eq:Ptheta}, evaluated from $\theta_{c}$ to $4$. Thus, the survival function $S(\rho_{th};\M,z)$ from eq. \eqref{eq:GW_numdens} is defined as:
\be
S(\rho_{th};\M,z) = \refrep{1} - T(\rho_{th}/\rho_{0})
\ee
where $T(\theta)$ is the integral of the angular orientation PDF $P(\theta)$.
Figure \ref{fig:theta} shows plots for these functions.

Combining these results, the expression for the number density of sources becomes
\be\label{eq:num_dens_merg}
n_{\rm GW}(z,\rho_{th}) =\tau \frac{R_{GW}(z)}{1+z}\int \d\M\ \phi(\M) \int_{\rho_{th}/\rho_0}^{4}{\rm d}\left(\frac{\rho}{\rho_0}\right)P(\rho/\rho_0) \quad [\text{Gpc}^{-3}] \,,
\ee
assuming an observation time $\tau$ and a redshift bin $z$. While the survival function is detector dependent, the chirp mass distribution depends on the mass distribution of black holes. To fully compute the biases we need to specify how likely are chirp masses of value $\M$.

The full derivation of the chirp mass PDF can be found in Appendix \ref{sec:chirp_calc}\footnote{We acknowledge and thank Michele Mancarella for help in this derivation.}. Here we report only the final result for the PDF $\phi(\M)$:
\be \label{eq:chirp_pdf_method2_full}
\phi(\M) = \int_{m_{1}>(4/27)^{1/5}\M} \d m_{1}\ f[x_{1}(\M,m_{1})]g(m_{1})\frac{\dd{x_{1}(\M,m_{1})}}{\dd{z}}\,,
\ee
where $f$ and $g$ are the distributions of secondary and primary masses respectively. $x_{1}$ is a real solution for the secondary mass $m_2$ in terms of chirp $\M$ and primary $m_{1}$ masses,
given by the expression
\bea
x_{1} = \frac{\M^{5/3}}{2^{1/3}3^{2/3}m_{1}^{3/2}}\Bigg[&\left(9m_{1}^{5/2}+\sqrt{81m_{1}^{5}-12\M^{5}}\right)^{1/3}&\nn\\
+&\left(9m_{1}^{5/2}-\sqrt{81m_{1}^{5}-12\M^{5}}\right)^{1/3}&\Bigg] \,.
\eea
Note that this expression is obtained by solving the cubic equation \eqref{eq:chirp_xy} and is valid for $m_1 > m_2$.

We then employ current phenomenological distributions of primary and secondary masses from the catalogues of the LIGO-Virgo-KAGRA (LVK) collaboration \cite{GWTC3,GWTC2}, to describe $f$ and $g$ in eq. \eqref{eq:chirp_pdf_method2_full}. Their full prescriptions can be found in Appendix \ref{sec:mass_distributions}.

\subsection{Magnification and evolution biases for GW events}\label{sec:gw_biases}
Using the definition of magnification bias described in eq. \eqref{eq:mag_bias}, we can write:
\bea\label{eq:mag_bias_expr}
    s^{GW}(z, \rho_{th}) &=& -\frac{1}{5}\frac{\partial\ln n(z,\rho_{th})}{\partial\ln\rho_{th}}\bigg|_z = -\frac{1}{5}\frac{\rho_{th}}{n(z,\rho_{th})}\frac{\partial n(z,\rho_{th})}{\partial\rho_{th}} \nn\\
    &=& -\frac{1}{5}\frac{\rho_{th}}{n(z,\rho_{th})}\left[-\tau\frac{R_{GW}(z)}{1+z}\int \d\M\ \phi(\M)P\left(\frac{\rho_{th}}{\rho_0}\right)\frac{1}{\rho_0}\right]\nn\\
    &=& \frac{\rho_{th}}{5}\frac{\int \d\M\ \phi(\M)P(\frac{\rho_{th}}{\rho_0})\frac{1}{\rho_0}}{\int \d\M\ \phi(\M)S(\rho_{th};\mathcal{M},z)}\,.
\eea
\begin{figure}
    \includegraphics[width=0.9\textwidth]{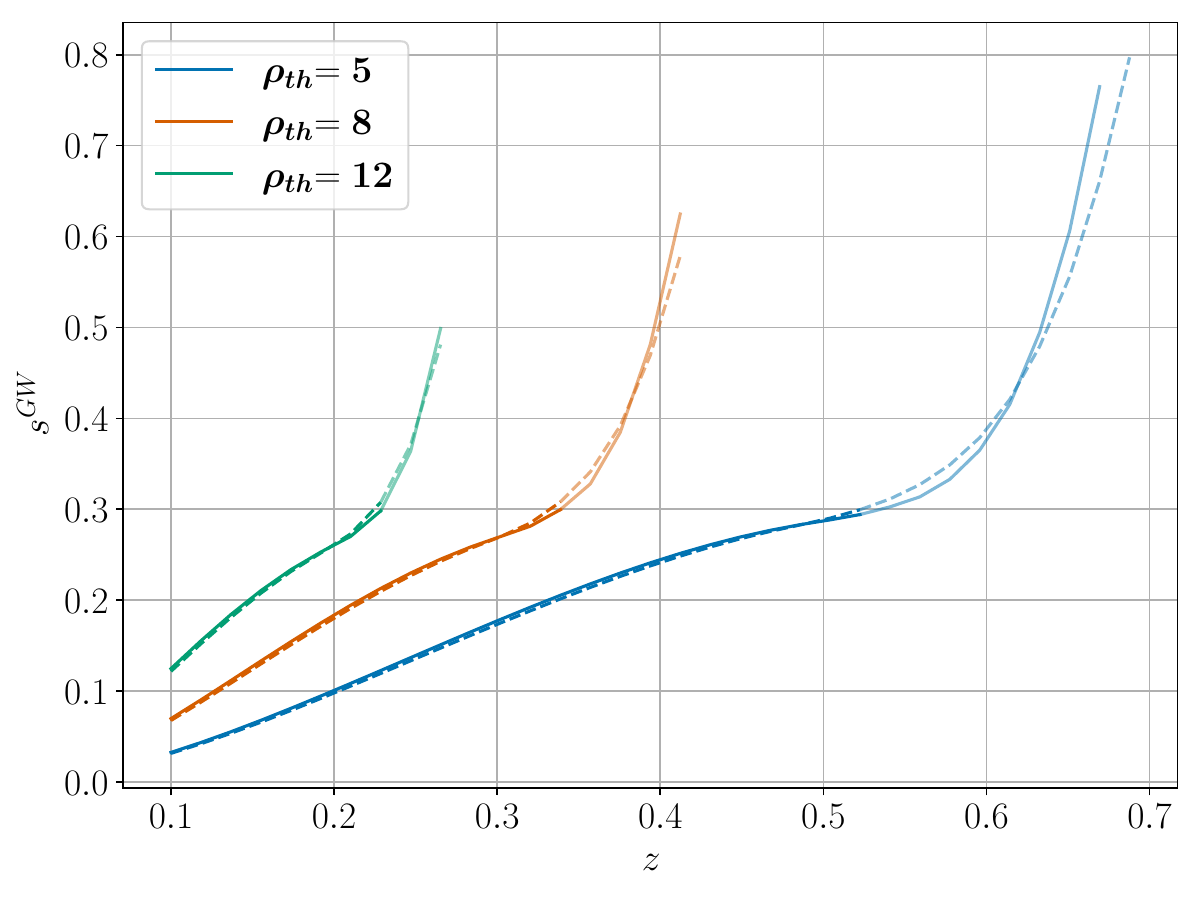}
    \caption{Magnification bias for an LVK-era experiment. We stop calculating $s$ for each model when the number of observed sources is below the arbitrary value of 100 events Gpc$^{-3}$; this is to suggest a limit of validity for the bias, as clustering analysis requires a larger number of sources. Solid lines indicate a Power Law + Peak model for the primary BH mass, whilst dashed ones show the Broken Power Law model. Fainter lines represent the range where the biases become more uncertain, as the number of observed sources crosses $120$, getting closer to our arbitrary cut.}
    \label{fig:mag_biases_ligo}
\end{figure}
Initially, we plug in the sensitivity curve for a network of aLIGO-like interferometers to mimic the LVK detectors\footnote{https://dcc.ligo.org/LIGO-T1800044/public}\refrep{, adding the SNR in quadrature for three detectors}.
We find that the values of the magnification bias in this case are strongly dependent on the low number of sources detectable. In fact, figure \ref{fig:mag_biases_ligo} shows very large values of $s$ for all models considered. Clearly, assuming a higher threshold of detection drastically decreases the number of observed sources even at lower redshift, and, consequently, increases the magnification bias. Conversely, a lower value of $\rho_{th}$ (e.g. blue line) yields lower values of $s$ across a larger redshift range. Further, it is interesting to note that, at the same detector's threshold, different distributions of chirp masses have similar magnification biases. 
\refrep{This suggests that for the chirp mass models assumed in this paper, $s$ does not vary significantly. We note that this conclusion is not necessarily true for any chirp mass distribution. }

\begin{figure}
    \centering
    \includegraphics[width=0.95\textwidth]{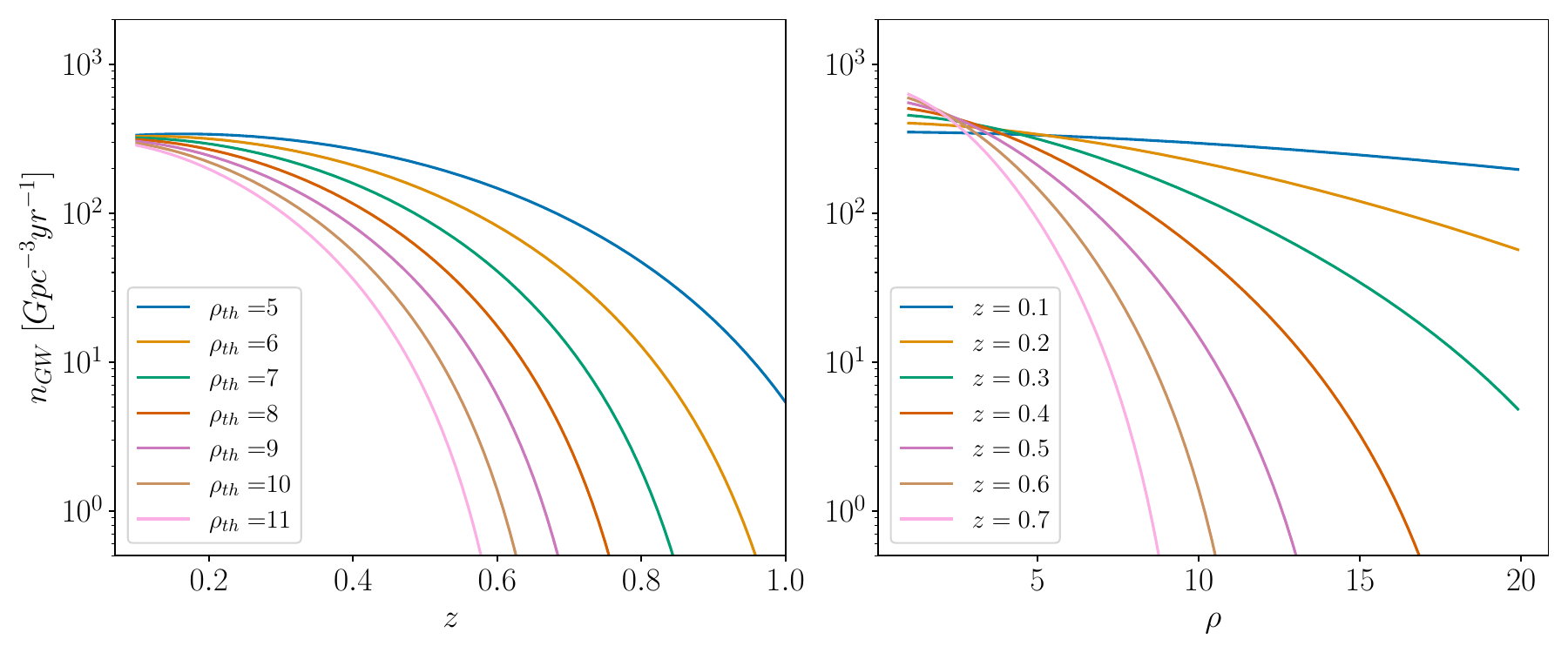}
    \caption{Modelling the number densities of GWs observable with LVK \refrep{using \eqref{eq:GW_numdens}}. {\em Left}: as a function of redshift for different (fixed) SNR threshold values; {\em Right}: as a function of $\rho_{th}$ at different (fixed) values of redshift. At each given $z$, the value of the magnification bias is given by the slope of the corresponding line in the right-hand plot. Increasing $\rho_{th}$ starts to change significantly around $z=0.5$, explaining the behaviour of $s$ in figure \ref{fig:mag_biases_ligo}.}
    \label{fig:n_for_rhos}
\end{figure}

The large values of $s$ for LVK shown in figure \ref{fig:mag_biases_ligo} can be further explained with figure \ref{fig:n_for_rhos}, investigating the dependence of the observed number density of GWs as seen by a LVK-like experiment on both redshift and SNR threshold.
\refrep{We calculate the observed number densities using \eqref{eq:num_dens_merg}, assuming an observation time of $1$ year and that the intrinsic merger rate $R_{GW}(z)$ follows directly the Madau-Dickinson rate \cite{YeFishback,MadauDickinson}:
\bea \label{eq:madau}
R_{GW} = R_0\frac{(1+z)^{2.7}}{1+(\frac{1+z}{2.9})^{5.6}} \, ,\
\eea
with $R_0$ providing the merger rate at $z=0$ and is given by \cite{GWTC2,GWTC3} as $R_0=23.9$ Gpc$^{-3}$yr$^{-1}$.} 
In particular, the left plot shows how a larger value of $\rho_{th}$ will yield a steeper curve at low redshift, and, conversely, a lower value of $\rho_{th}$ produces a more gentle one. Therefore, as redshift increases, so does the distance between each curve, showing the number of sources missed/detected when changing the minimum SNR. This is made more explicit in the plot on the right, where curves at fixed redshifts are plotted against $\rho_{th}$. Notably, the magnification bias can be read off directly here, as it is the slope at a given value of SNR threshold for a fixed redshift. Thus, one can see that approaching higher thresholds and redshifts, the slope increases drastically; in particular taking $\rho_{th}=8$ when going from $z=0.3$ to $z=0.4$ the slope is evidently steeper, which results in the much steeper values of the bias in figure \ref{fig:mag_biases_ligo}. 

\begin{figure}
    \includegraphics[width=\textwidth]{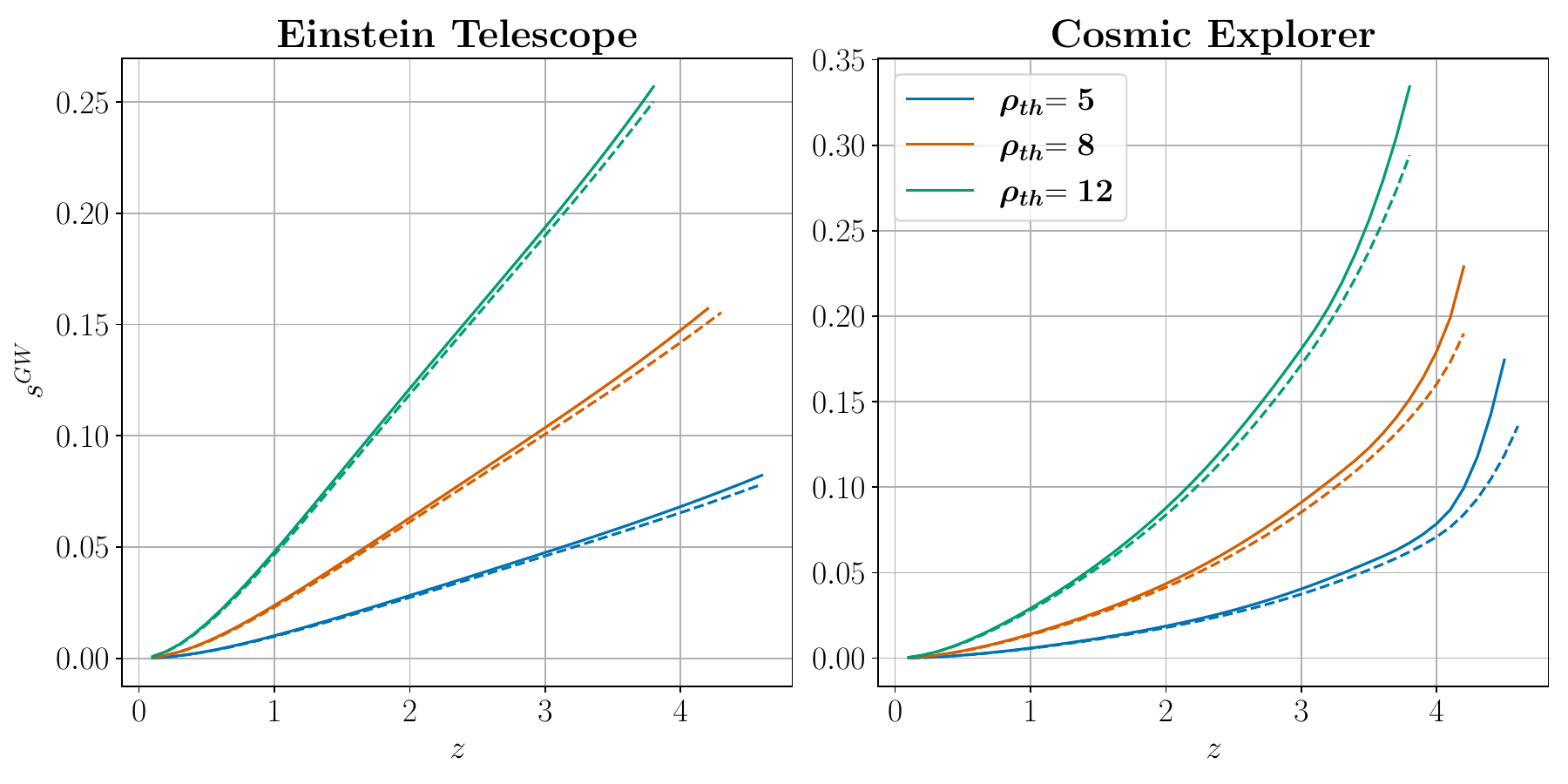}
    \caption{Magnification bias for third-generation detectors, {\em left} ET, {\em right} CE. As before, solid lines are for a Power Law + Peak distribution of the primary BH mass, and dashed ones are for the Broken Power Law model.}
    \label{fig:mag_biases_3G}
\end{figure}

We then plot in figure \ref{fig:mag_biases_3G} the magnification bias for third-generation detectors ET and CE. The increased sensitivity of these future experiments ensures that almost all sources are observed, thus the bias will be extremely small. The two detectors differ only towards higher redshifts, with CE showing a steeper tail similar to the LVK case. This could be explained by the different sensitivity curves. CE is sensitive from $5$Hz, whereas ET from $1$Hz. However, GWs emitted at larger distance will result in a lower merging frequency shown in eq. \eqref{eq:fmax}, thus a smaller frequency ranges observable by the detector. Therefore, at higher distances, this might affect the observed number of sources and give rise to the difference seen in the two biases in \ref{fig:mag_biases_3G}. 

For further convenience, we provide functional forms for the magnification biases calculated so far. We fit a polynomial of order $3$ with coefficients $y = a+bx+cx^2+dx^3$, 
and note the fitting values in table \ref{tab:s_funcs}. Additionally, we opted to fit only the initial part of the curve for LVK, thus before the sudden steepening. As the curves all steepen when the number of observed sources is $n_{obs}\sim120$, we chose this value as upper cut for the functional forms. The appropriate redshift intervals are listed in the final column in table \ref{tab:s_funcs}.

\begin{table}
    \centering
    \begin{tabular}{|c|c|c|c|c|c|c|c|}   
    \hline
    & $\rho_{th}$ & $a$ & $b$ & $c$ & $d$ & $z$\\
    \hline
    \multirow{3}{*}{ET} &
    $5$ & \num{-1.47e-03} & \num{7.38e-03}&\num{5.06e-03}&\num{-6.84e-04} & \multirow{3}{*}{$[0.1,3.5]$}\\
    \cline{2-6} &
    $8$ & \num{-3.77e-03} & \num{1.97e-02}&\num{9.43e-03}&\num{-1.34e-03} &\\
    \cline{2-6} &
    $12$ & \num{-8.39e-03} & \num{4.54e-02}&\num{1.36e-02}&\num{-2.04e-03} &\\
    \hline
    \multirow{3}{*}{CE} & $5$ &
    \num{-1.06e-03} & \num{5.39e-03}&\num{1.06e-03}&\num{5.89e-04} & \multirow{3}{*}{$[0.1,3.5]$}\\
    \cline{2-6} & $8$
     & \num{-1.87e-03} & \num{1.06e-02}&\num{4.56e-03}&\num{7.36e-04} &\\
    \cline{2-6} & $12$
     & \num{-5.59e-03} & \num{2.92e-02}&\num{3.44e-03}&\num{2.58e-03} &\\
     \hline
     \multirow{3}{*}{LVK} & $5$ &
     \num{-2.35e-02} & \num{4.14e-01}&\num{1.47e+00}&\num{-2.10e+00} & $[0.1,55]$\\
    \cline{2-7} & $8$
     & \num{-8.07e-02} & \num{1.63e+00}&\num{-1.57e+00}&\num{1.79e-01} & $[0.1,0.35]$\\
    \cline{2-7} & $12$
     & \num{-1.22e-01} & \num{3.15e+00}&\num{-7.61e+00}&\num{7.33e+00} & $[0.1,0.22]$\\
     \hline
    \end{tabular}
    \caption{Coefficients of a third-order polynomial fit to the magnification bias for GWs detectors for different SNR threshold.}
    \label{tab:s_funcs}
\end{table}

\begin{figure}
    \centering
    \includegraphics[width=\textwidth]{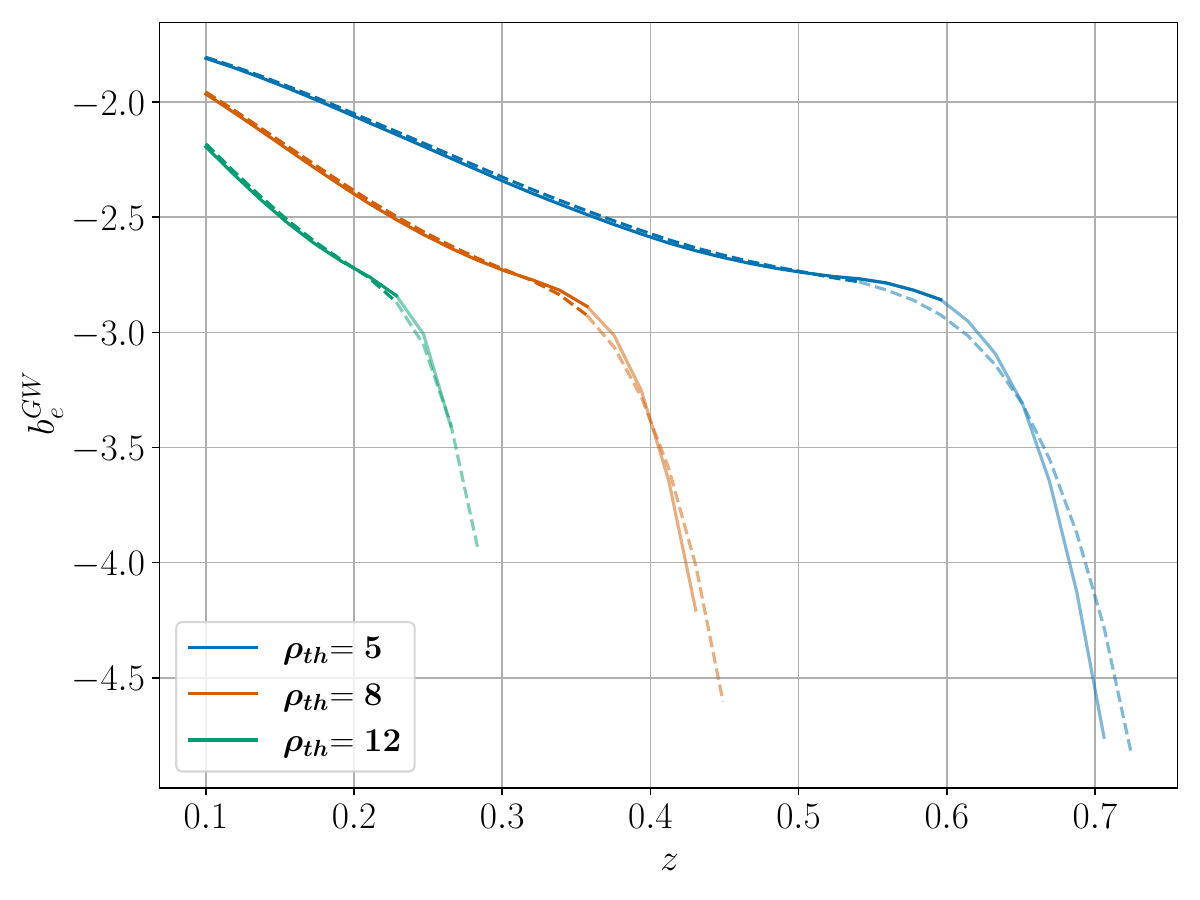}
    \caption{Evolution bias for a LVK-like survey. As previously, solid lines indicate a Power Law + Peak model for the primary BH mass, dashed for the Broken Power Law model. Similarly to figure \ref{fig:mag_biases_ligo}, faint lines represent a regime with observed number of sources $100<n_{obs}<120$, thus approaching our arbitrary cut for the biases.}
    \label{fig:evo_bias_LIGO}
\end{figure}

\begin{figure}
    \centering
    \includegraphics[width=\textwidth]{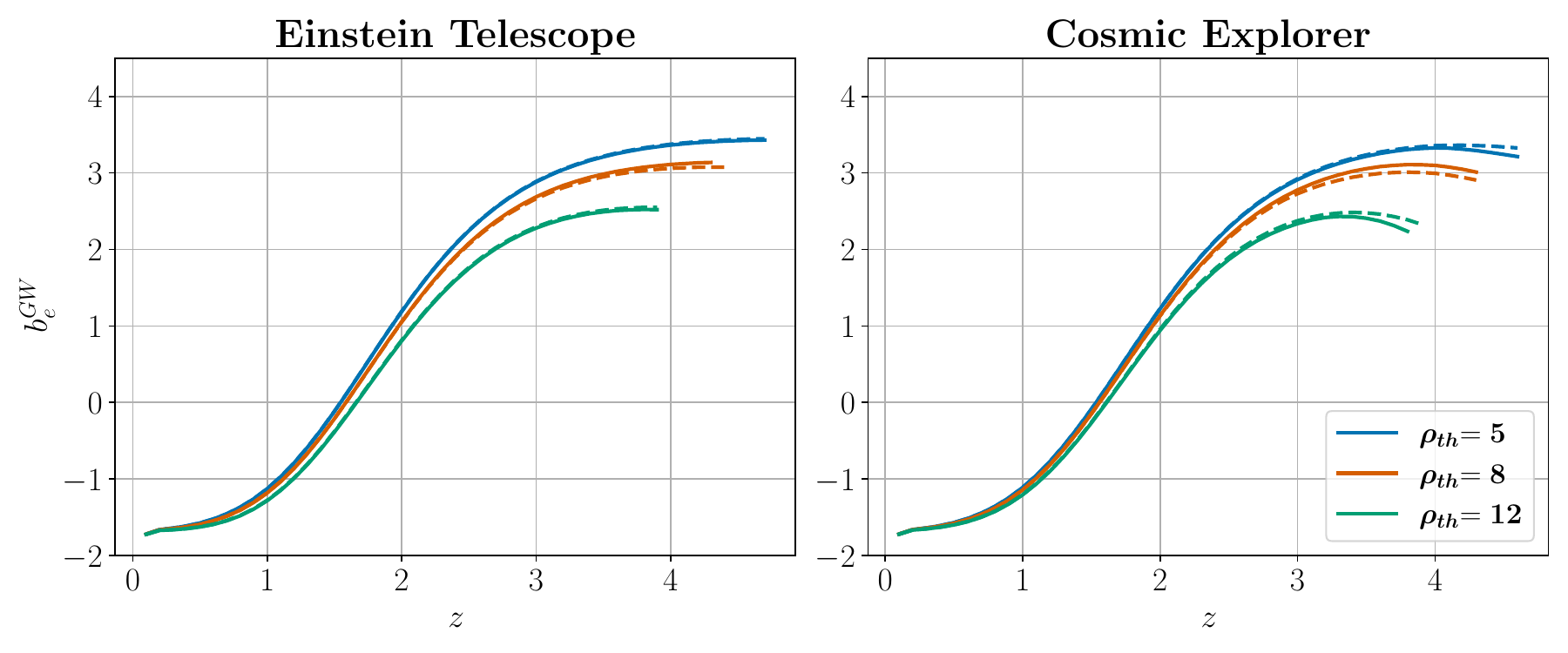}
    \caption{Evolution bias for third-generation ground based GWs observatories ET ({\em left}) and CE ({\em right}). These will be able to trace the evolution of BBH mergers up to high redshift.}
    \label{fig:evo_bias_3G}
\end{figure}

The evolution bias follows similarly.
Using its definition in eq. \eqref{eq:be_def}, we can find:
\bea
    b_{e}^{GW}(z,\rho_{th}) &=& \frac{\partial\ln n(z,\rho_{th})}{\partial\ln a}\bigg|_{\rho_{th}} = -\frac{1+z}{n(z,\rho_{th})}\frac{\partial n(z,\rho_{th})}{\partial z}\nn\\
    &=& -\frac{1+z}{n(z,\rho_{th})} \tau\bigg\{\frac{\partial}{\partial z}\left(\frac{R_{GW}(z)}{1+z}\right)\int \d \M\ \phi(\M)S(\rho_{th};\mathcal{M},z)+ \nn\\
    && + \frac{R_{GW}(z)}{1+z}\int \d\M \ \phi(\M)\left[P\left(\frac{\rho_{th}}{\rho_0}\right)\left(\frac{\rho_{th}}{\rho_0^2}\right)\frac{\d\rho_0}{\d z}\right]\bigg\}
\eea
We plot $b_e$ for a LVK-like experiment in figure \ref{fig:evo_bias_LIGO}, and for third-generation detectors ET and CE in figure \ref{fig:evo_bias_3G}. For the former, the lower sensitivity results in a lower number of detections and thus a much stronger correction. In fact, a strongly negative evolution bias implies that the survey observes a population which appears to become more sparsely distributed as the universe evolves. However, this is actually an effect of the low sensitivity of the experiment, which results in a lower number of detections, and thus it appears as if the sources are intrinsically decreasing with redshift, whilst we should expect a larger volume of sources as we approach cosmic dawn. In fact, the improved sensitivity of 3G detectors ET and CE results in correctly tracing the evolution of BBH mergers. Considering a Madau-Dickinson rate peaking around cosmic dawn as in eq. \eqref{eq:madau}, i.e. $z_c\sim2$, the evolution bias shows a population of tracers becoming more densely distributed as the universe expands, until it reaches $z=2$, where the trend inverses; then, as the tracers start to become more sparsely populated as the universe expands (i.e. the scale factor grows), the evolution bias becomes negative towards present time. 

Finally, we fit cubic polynomials to the models for $b_e^{GWs}$ calculated so far. As with the magnification bias for LVK, we only fit curves in redshifts for which $n_{obs}>120$, which we report in the final column of table \ref{tab:evo_funcs}. This was to avoid the final part of the curves, which are drastically impacted by lower observations and thus more uncertain.

\begin{table}
    \centering
    \begin{tabular}{|c|c|c|c|c|c|c|}   
    \hline
    & $\rho_{th}$ & $a$ & $b$ & $c$ & $d$ & $z$\\
    \hline
    \multirow{3}{*}{ET} &
    $5$ & \num{-1.48e+00} & \num{-1.23e+00}&\num{2.01e+00}&\num{-3.69e-01} & \multirow{3}{*}{$[0.1,3.5]$} \\
    \cline{2-6} &
    $8$ & \num{-1.46e+00} & \num{-1.29e+00}&\num{2.00e+00}&\num{-3.66e-01} &\\
    \cline{2-6} &
    $12$ & \num{-1.45e+00} & \num{-1.39e+00}&\num{1.98e+00}&\num{-3.63e-01} &\\
     \hline
    \multirow{3}{*}{LVK} &
    $5$ & \num{-1.50e+00} & \num{-3.52e+00}&\num{1.13e+01}&\num{-3.11e+01} & $[0.1,0.55]$ \\
    \cline{2-6} &
    $8$ & \num{-1.34e+00} & \num{-6.89e+00}&\num{7.73e+00}&\num{-2.10e+00} & $[0.1,0.35]$\\
    \cline{2-6} &
    $12$ & \num{-1.04e+00} & \num{-1.76e+01}&\num{1.05e+02}&\num{-4.36e+02} & $[0.1,0.22]$\\
     \hline
    \end{tabular}
    \caption{Coefficients of a third-order polynomial fit to the evolution bias for GWs detectors for different SNR threshold. Given the negligible difference in the biases between ET and CE over the redshift interval $z \in [0.1,3.5]$, we only report the fits to ET.}
    \label{tab:evo_funcs}
\end{table}

\section{Supernovae biases}\label{sec:SNIa}
\begin{figure}
    \centering
    \includegraphics[width=\textwidth]{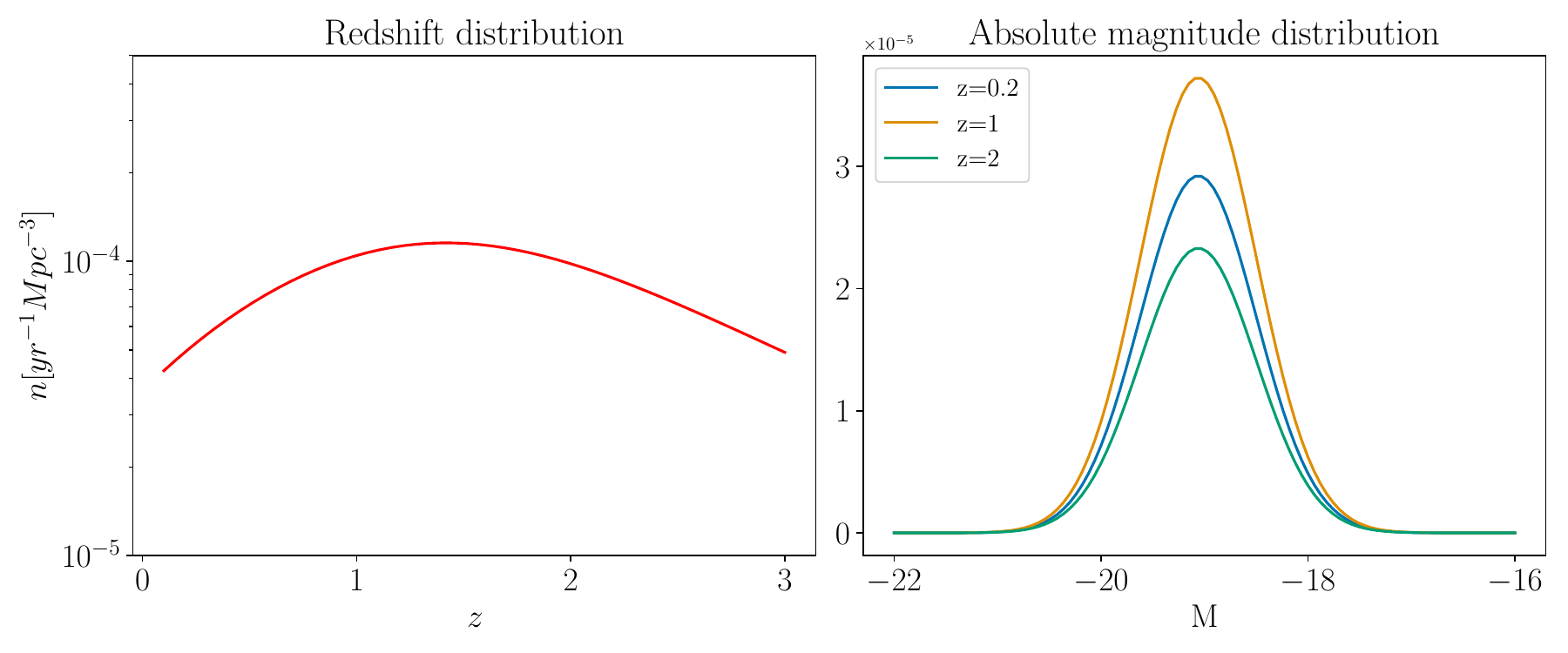}
    \caption{{\em Left:} Redshift distribution of SNIa, $n_{SNIa}$, as prescribed from eq. \eqref{eq:intrinsic_supernovae}. The logarithmic scale shows little variation of the rate across the redshift range, peaking some time after cosmic noon as expected. {\em Right:} Absolute magnitude distribution of SNIa at different redshift. The peak follows closely the $z$ distribution, as this is simply a Gaussian modulated by $n_{SNIa}$.
    }
    \label{fig:SNae_distr}
\end{figure}
Once detected, SN1a can be followed up with further observations in different bands, or spectroscopically. However, the detection of a supernova is fundamentally dependent on its observed magnitude or luminosity. It follows that the magnification bias for a sample of SNIa should be defined in the same manner as for galaxy samples, i.e.:
\be\label{eq:mag_def_sneia}
s^{SNIa}\equiv \frac{\partial\log_{10} n^{SNIa}}{\partial m}\bigg|_z \,.
\ee 
Therefore one only needs to model the number density of SNIa as a function of time and magnitude threshold.

\subsection{SNIa event rate}

In order to model the number density of SNIa, we employ a parameterisation described in \cite{Oguri_2010}, constructing a luminosity function $\Phi(M,z)$, with $M$ being the peak absolute magnitude of the event. A key assumption in this analysis is that $M$ is given by a narrow Gaussian distribution \cite{Oda_2005, Yasuda_2010}.

The starting point of the expression is the star formation rate ($SFR$). Initially we pick a standard cosmic star formation rate produced \cite{Hopkins_2006, Madau_2014}:

\bea \label{eq:sfr}
SFR(z)=\frac{(0.0118+0.082z)h}{1+(z/3.3)^{5.2}} 
M_{\odot}\, \text{yr}^{-1}\, \text{Mpc}^{-3}\, .\
\eea
Further, the delay time between formation of a binary and subsequent SNIa explosion is modelled by \cite{Totani_2008} as a power law distribution:
\be \label{eq:td_dist}
f(t_{D}) \propto t_{D}^{-1.08} \quad (t_{D}>0.1 \textrm{Gyr}) \, .\
\ee
Combining the above, we have the SNIa rate in units of yr$^{-1}$Mpc$^{-3}$ \cite{Oguri_2010}:
\be \label{eq:intrinsic_supernovae}
n_{SNIa}(z) = \eta\, C_{SNIa} \frac{\int_{0.1}^{t(z)} SFR[z(t-t_D)]f(t_D) \d t_D\ }{\int_{0.1}^{t(z=0)}f(t_D)\d t_D\ } \, .\
\ee
where the factor $C_{SNIa}=0.032M_{\odot}^{-1}$ can be computed from the stellar mass range of $3M_{\odot}<M<8M_{\odot}$ for SNIa and the initial mass function \cite{Baldry_2003}. Additionally, the explosion efficiency $\eta$ is taken as the canonical value of $0.04$ \cite{Hopkins_2006}. In the left panel of figure \ref{fig:SNae_distr} one can see the flat rate density of SNIa as a function of redshift. 

Finally, assuming the absolute magnitudes of supernovae are Gaussian-distributed, we can produce the magnitude distribution of sources \cite{Oguri_2010}:
\be \label{eq:luminosity_dist_supernovae}
\frac{d\Phi(M,z)}{dM} = \frac{n_{SNeIa}(z)}{1+z}\frac{1}{\sqrt{2\pi}\sigma}\exp{\left[-\frac{(M-M^{\ast})^2}{2\sigma^2}\right]}=\frac{n_{SNeIa}(z)}{1+z}G(M^{\ast},\sigma) \, ,\
\ee
where $G$ is a Gaussian distribution, $M^{\ast}=-19.06$ in the B-band and $\sigma = 0.56$. $\Phi(M,z)$ is the magnitude-equivalent of a luminosity function for SNIa. In the right panel of figure \ref{fig:SNae_distr} one can see the distribution for different redshifts. Here we assume that the dispersion of the distribution is the same remains constant.

The limiting absolute magnitude of detection is found using the limiting apparent magnitude of the detector in a specific band, the redshift of the event and the cross-filter correction $K$. Recall that at peak brightness:
\be \label{eq:magnitude}
m = M +5\log_{10}(D_L[Mpc])+25 \, ,\
\ee
with $m$ and $M$ being, respectively, the apparent and absolute magnitude, and where we assume that the single or cross-filter K-correction is simply a constant offset \refrep{(note that we neglect errors on the measured apparent magnitude)}.  
Thus, the number of sources that have an absolute magnitude smaller than the limiting value - and so are bright enough to be observed - is just the integral of the luminosity function over the allowed magnitudes:
\be \label{eq:n_supernovae}
n(M<M_{max},z) = \int_{-\infty}^{M_{max}} \d M\ \frac{d\Phi(M,z)}{dM} \,.\
\ee

\subsection{Magnification and evolution biases of SNIa}\label{sec:sneia_biases}
\begin{figure}
    \centering
    \includegraphics[width=\textwidth]{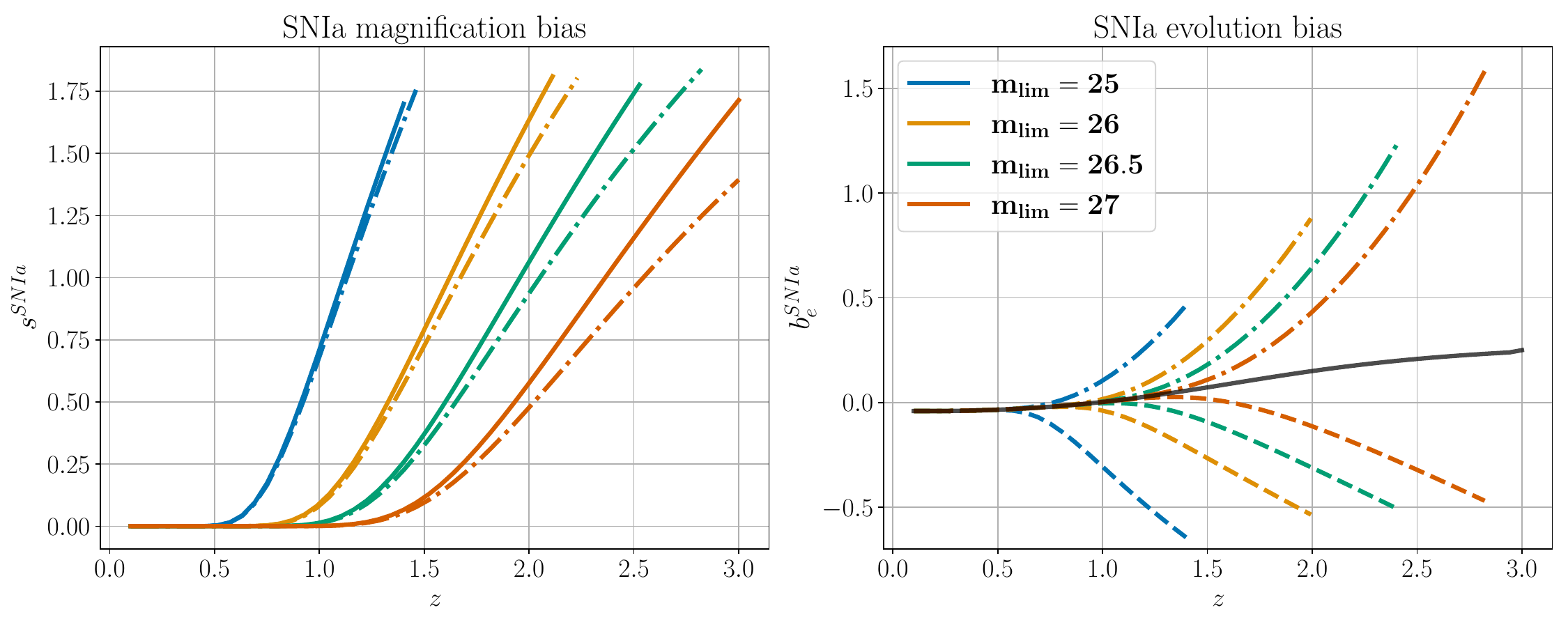}
    \caption{
    {\em Left}: Magnification biases for a SNIa survey. Solid lines represent the biases for a SNIa with fixed luminosity, whilst dash-dot lines illustrates the bias for SNIa with an intrinsic luminosity redshift dependence.
    {\em Right} Evolution biases for a SNIa survey. Coloured lines show models for an intrinsic SNIa luminosity evolving with time, whilst the black line describe SNIa of fixed peak magnitude $M_{peak}=-19.06$. In particular, dashdot lines represent an evolution with a power law of index $\delta_2=2$, whilst dashed ones have $\delta_1=0.29$. For the magnification bias, the two overlap almost completely, whilst this is not the case for the evolution bias, due to extra redshift dependence. \refrep{Note that the legend is shared between the two plots.} }
    \label{fig:biases_SNIa}
\end{figure}

As the magnification bias is computed at a fixed redshift, the K-correction can be safely considered constant, thus neglected when differentiating. Thus, the magnification bias can be evaluated by using \eqref{eq:mag_def_sneia}:
\be\label{eq:mag_bias_magnitude}
s = \frac{\partial \log_{10} n(m_{max},z)}{\partial m_{max}}\bigg|_{z}= \frac{\partial \log_{10} n(M_{max},z)}{\partial M_{max}}\bigg|_{z} \, .\
\ee
Finally, we find:
\bea
s_{SNIa} &=& \frac{1}{n\ln10}\frac{\partial n}{\partial M} = \frac{1}{\ln10}\frac{G(M_{max},\sigma)}{\int_{-\infty}^{M_{max}}\d M\ G(M_{max},\sigma)}\nn\\
&=&\frac1{\ln10}\frac{G(M_{max},\sigma)}{\frac1{2}\left(1+\textrm{erf}\left[{\frac{M-M^{\ast}}{\sqrt{2\pi}\sigma}}\right]\right)} \,.
\eea
Thus, using eq. \eqref{eq:magnitude}, we can compute the limiting absolute magnitude at each given redshift from a value of the limiting apparent magnitude of a telescope and compute the magnification bias as prescribed in eq. \eqref{eq:mag_bias_magnitude} (see figure \ref{fig:biases_SNIa}).

The evolution bias for SNIa is computed similarly, from eq. \eqref{eq:be_def}:
\bea \label{eq:evo_SNIa}
b_e^{SNIa} &=& \frac{\partial\ln n}{\partial\ln a}\bigg|_{M_{max}} = -\frac{1+z}{n(z,M_{max})}\int_{-\infty}^{M_{max}(z)} \d M\ \frac{\partial}{\partial z}\frac{\d\Phi(M,z)}{\d M} \, .\
\eea
Both biases are computed up to a value of redshift for which we can observe less than $2\sigma$ of the magnitude distribution; this was an arbitrary cut to select only the portion of observations which would be statistically significant.

From figure \ref{fig:biases_SNIa} we note that the magnification bias is fixed to zero until a value of redshift dependent on the magnitude cut. This is due to the fact that before such distance, all SNIa can be observed; however, at such redshift the limiting absolute magnitude approaches the Gaussian distribution of SNIa, thus increasing the number of objects near the threshold (and consequently, of objects being magnified in or out).

As we did for the GWs biases, we fit a cubic polynomial of equation $y=a+bx+cx^2+dx^3$ to the values of $s^{SNIa}$ calculated here, and we report them in table \ref{tab:SNIa_s_funcs}.

\begin{table}
    \centering
    \begin{tabular}{|c|c|c|c|c|c|}   
    \hline
    $m_{lim}$ & $a$ & $b$ & $c$ & $d$ & $z$\\
    \hline
    $25$ & \num{1.72e+00} & \num{-7.12e+00}&\num{8.69e+00}&\num{-2.59e+00} & $[0.5,1.4]$ \\
    \hline
    $26$ & \num{1.35e+00} & \num{-4.20e+00}&\num{3.68e+00}&\num{-7.59e-01} & $[1,2]$\\
    \hline 
    $26.5$ & \num{1.66e+00} & \num{-3.95e+00}&\num{2.77e+00}&\num{-4.71e-01} & $[1,2.5]$\\
    \hline
    $27$ & \num{1.62e+00} & \num{-3.22e+00}&\num{1.87e+00}&\num{-2.64e-01} & $[1,3]$\\
    \hline
    \end{tabular}
    \caption{Coefficients of a third-order polynomial fit to the magnification bias for SNIa surveys with different limiting magnitudes. We report the redshift interval in the final column. Note, the bias is $0$ below the redshift range reported for each model.}
    \label{tab:SNIa_s_funcs}
\end{table}

If we then relax the assumption that SNIa have a fixed brightness across all redshifts \cite{Martinelli_2019, Linden_2009,Tutusaus_2017,Tutusaus_2018, Yijung_2020, Benisty_2022} then we can modify \eqref{eq:magnitude} and add an extra nuisance term $\Delta m^{evo}(z)$ on the right-hand side, accounting for a potential evolution of the SNIa intrinsic luminosity with redshift. Different models have been proposed to model this evolution, however we will consider only  Model B from \cite{Linden_2009}, also illustrated in \cite{Riess_2018}:
\be\label{eq:m_evo}
\Delta m^{evo}(z) = \epsilon z^\delta \, .
\ee
We can explore two sets of parameters for this, still in agreement with standard $\Lambda$CDM \cite{Tutusaus_2018}, i.e. $\epsilon_1 = 0.013\pm0.06$ and $\delta_1 = 0.29\pm0.22$, and $\epsilon_2 = 0.029\pm0.052$ and $\delta_2 = 2\pm1.7$. Both imply a population of SNIa which becomes intrinsically dimmer as the redshift increases.
The biases are then calculated as before, and then are plotted in figure \ref{fig:biases_SNIa}.

If the intrinsic luminosity of SNIa is allowed to vary with redshift, the effect is particularly relevant for the evolution bias, whilst the magnification one is only slightly affected. The reason is due to the nature of the luminosity function $\frac{\d\Phi}{\d M}$ in \eqref{eq:luminosity_dist_supernovae}. For a fixed luminosity kind of SNIa, the magnitude distribution function is separable in redshift and absolute magnitude; hence computing its evolution bias will result in simplifying out the integral over a Gaussian in magnitude (in eq. \eqref{eq:evo_SNIa}) and only terms related to the intrinsic redshift distribution of SNIa, and its derivative, will remain. These have a small contribution as the redshift distribution is assumed to be relatively flat, as seen in figure \ref{fig:SNae_distr}. Notably, the magnitude cut disappears and thus the evolution bias becomes independent of it. 

However, relaxing the assumption of a fixed luminosity brings in an extra redshift derivative of the Gaussian, as now the peak magnitude is dependent on redshift. This allows for the evolution bias to then depend on the magnitude cut as the integral is not simplified out anymore. 

Finally, we fit a cubic polynomial to these models for the evolution bias, reporting the coefficients of the functional forms in table \ref{tab:SNIa_evo_funcs}. 
\begin{table}
    \centering
    \begin{tabular}{|c|c|c|c|c|c|c|}   
    \hline
    & $m_{lim}$ & $a$ & $b$ & $c$ & $d$ & $z$\\
    \hline
    Fixed & / & \num{-3.07e-02} & \num{-7.24e-02}&\num{1.34e-01}&\num{-2.67e-02} & $[0.1,3]$ \\
    \hline
    \multirow{4}{*}{$\delta_1$} &
    $25$ & \num{-1.41e-01} & \num{7.05e-01}&\num{-1.07e+00}&\num{2.11e-01} & $[0.5,1.4]$ \\
    \cline{2-7}
    & $26$ & \num{-9.08e-02} & \num{2.49e-01}&\num{-1.86e-01}&\num{-3.32e-02} & $[1,2]$\\
    \cline{2-7} 
    & $26.5$ & \num{-9.12e-02} & \num{2.09e-01}&\num{-1.09e-01}&\num{-2.30e-02} & $[1,2.5]$\\
    \cline{2-7}
    & $27$ & \num{-7.75e-02} & \num{1.11e-01}&\num{1.51e-02}&\num{-3.87e-02} & $[1,3]$\\
    \hline
    \multirow{4}{*}{$\delta_2$} &
    $25$ & \num{-3.78e-02} & \num{1.48e-02}&\num{-1.84e-01}&\num{3.12e-01} & $[0.5,1.4]$ \\
    \cline{2-7}
    & $26$ & \num{-4.69e-02} & \num{7.66e-02}&\num{-2.21e-01}&\num{2.13e-01} & $[1,2]$\\
    \cline{2-7} 
    & $26.5$ & \num{-4.09e-02} & \num{4.39e-02}&\num{-1.47e-01}&\num{1.47e-01} & $[1,2.5]$\\
    \cline{2-7}
    & $27$ & \num{-5.24e-02} & \num{9.59e-02}&\num{-1.72e-01}&\num{1.23e-01} & $[1,3]$\\
    \hline
    \end{tabular}
    \caption{Coefficients of a third-order polynomial fit to the evolution bias for SNIa surveys with different limiting magnitudes. As before, We report the mean squared error (MSE) and redshift interval in the final two columns. Note, the first model refers to a population of SNIa of fixed intrinsic luminosity, and yields a result independent of the limiting magnitude. Models $\delta_1$ and $\delta_2$ describe SNIa with intrinsic luminosity evolving with redshift at different rates.}
    \label{tab:SNIa_evo_funcs}
\end{table}

\section{Impact on observables}\label{sec:applications}

In this section, we will show the relevance of these biases in the luminosity distance clustering power spectra in two ways. 
Initially, we will investigate their impact on the relativistic corrections to the number counts which are most likely to be detected, namely the Doppler and lensing terms. Then, we will examine how these biases affect the angular power spectrum, exploring both auto-correlation and cross-bin correlations at different redshifts.

\subsection{Impact on the number counts}
The modelling of these biases is extremely important to investigate the relativistic corrections in the number density fluctuation in eq. \eqref{eq:general_num_contrast}. As shown in \cite{Us}, when analysing clustering in luminosity distance space several terms are dependent on these two parameters, such as the lensing and Doppler terms. Notably, the lensing term in luminosity distance is dependent on both bias parameters, as opposed to the redshift-space case which depends only on the magnification bias.

In particular, by computing the appropriate perturbation in luminosity distance and isolating each correction to the underlying matter density $\delta_n$, the number density fluctuation in eq. \eqref{eq:general_num_contrast} is recast into \cite{Us}:
\bea\label{eq:coefficients}
\Delta(\boldsymbol{n},D_L) &=& \delta_n+A_D(\bm v\cdot\bm n)+A_{LSD}\partial_r(\bm v\cdot\bm n) +\frac1{\bar r}\int_0^{\bar r} \d r\ A_{L}\Delta_{\Omega}(\Phi+\Psi) +\cdots \, \
\eea
where we only report the main correction terms, with coefficients
\bea
&A_D& = 1-2(\gamma+\beta) \label{eq:AD}\, ,\\
&A_{LSD}& = -2\ \frac{\gamma}{\mathcal{H}} \label{eq:ALSD}\, ,\\
&A_{L}& = \frac12 \left[\left(\frac{\bar r-r}{r}\right)(\beta-2)+\frac{1}{1+\bar r \mathcal{H}}\right]\,. \label{eq:AL} 
\eea
Here $\bar{r}$ is the source's position, $r$ the comoving distance, 
$\gamma \equiv \bar{r}\mathcal{H}/(1+\bar{r}\mathcal{H})$, and where we defined
\bea
\beta&\equiv&
\refrep{ 1}-5s+\gamma\left[\frac2{\bar r \mathcal{H}}+\gamma\left(\frac{\cal{H}'}{\mathcal{H}^2}-\frac1{\bar r\cal{H}}\right)-1-b_e\right] \, .
\eea
\begin{figure}
    \centering
    \includegraphics[width=\textwidth]{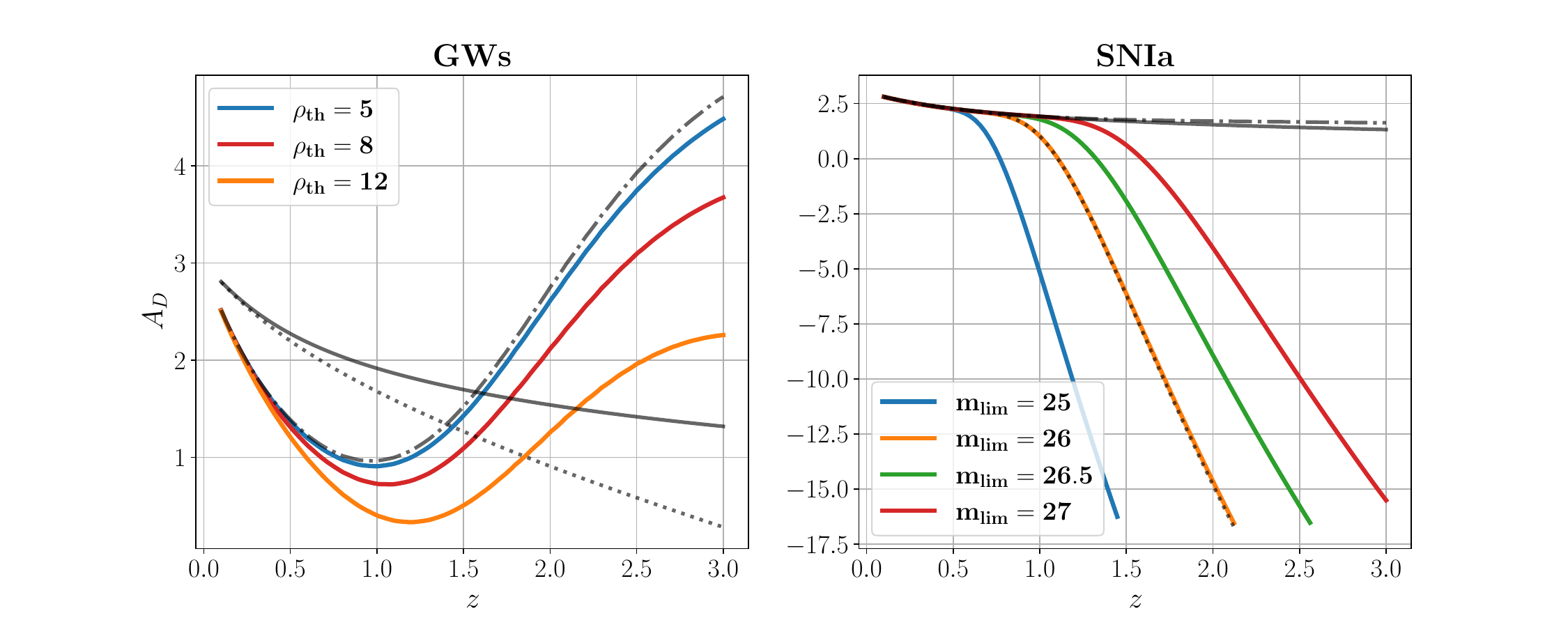}
    \caption{\refrep{Dimensionless} Doppler kernel $A_D$ including the contribution of the biases. {\em Left}: the effect when considering GWs observed by ET (given the similarities of the biases between ET and CE we plot only the former); {\em Right}: the effects when considering SNIa with different magnitude cuts. For both cases, we plot in grey the contribution in the case of the biases set to zero (solid line), only non-zero magnification bias (dashed) and only non-zero evolution bias (dash-dot); in particular, the latter two are set for $\rho_{th}=8$ on the left, and $m_{lim}=26$ on the right.
    }
    \label{fig:doppler}
\end{figure}
Note that $\beta$ contains the magnification bias $s$ and evolution bias $b_e$ of the source type in question. Thus, we explore the impact of the biases we calculated in Sections \ref{sec:gw_biases} and \ref{sec:SNIa} on the Doppler and lensing terms. 
We plot the Doppler amplitude $A_D$ 
in figure \ref{fig:doppler}, comparing the results for GWs (for an ET-like experiment) on the left panel and SNIa on the right one. We opt to show only one model of the biases for each population (i.e. Power Law + Peak chirp mass distribution for GWs, and SNIa with fixed intrinsic luminosity) for simplicity. Furthermore, the difference in the Doppler correction with the other models was found to be negligible. 
Additionally, we plot in grey the Doppler term with, respectively, the biases set to zero (solid line), only magnification (dashed) and only evolution (dash-dot). 
This clearly shows that the biases dominate the amplitude of the Doppler correction and are crucial for its analysis. In particular, in the case of GWs from an ET-like experiment, the correction traces the evolution bias significantly, whilst for SNIa, the higher values of magnification bias impact the Doppler term and become the main contribution. 

\begin{figure}
    \centering
    \includegraphics[width=\textwidth]{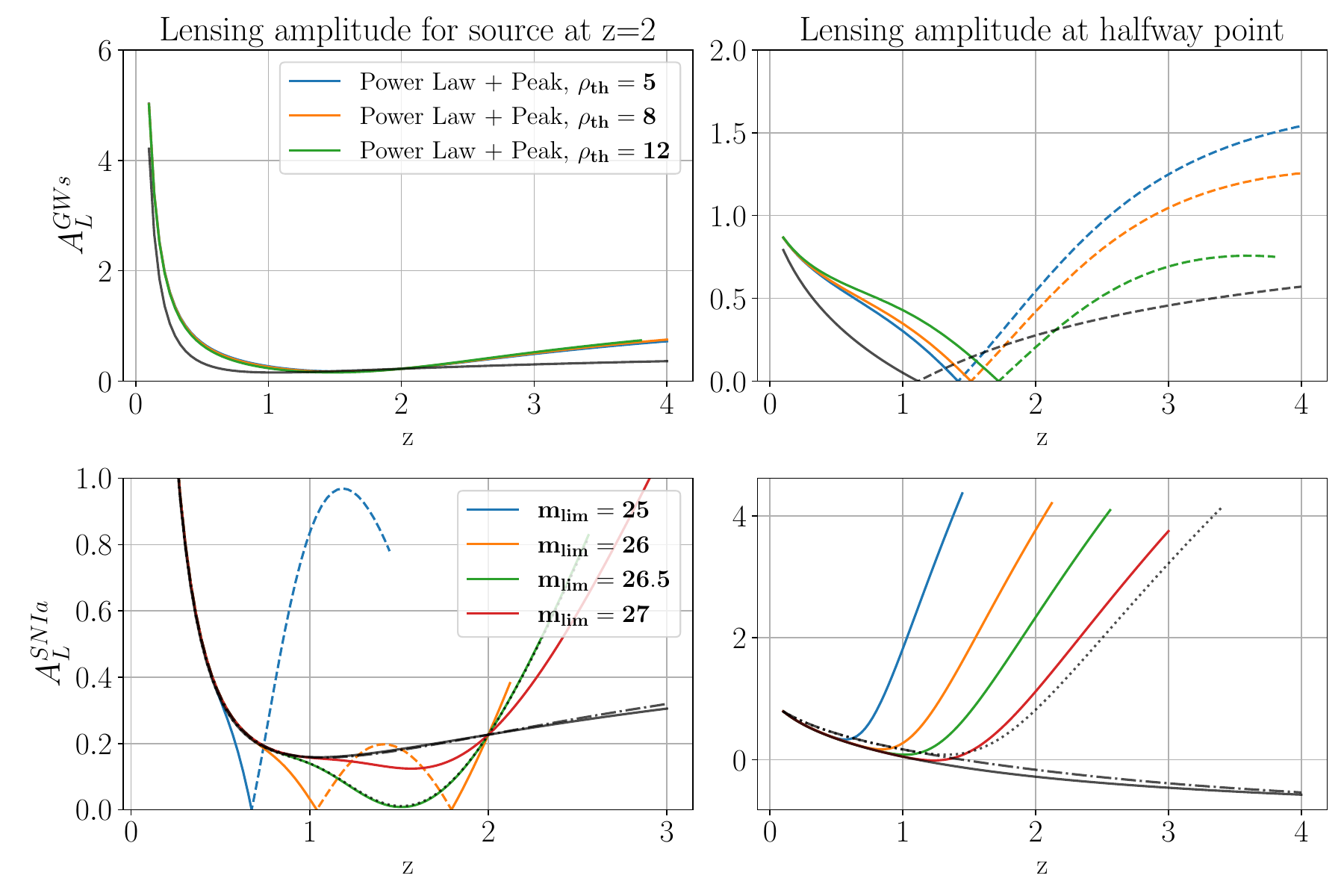}

    \caption{Comparison between the lensing amplitude $A_L$ for an ET-like experiment detecting GWs assuming different SNR-thresholds $\rho_{th}$ ({\em top panels}) and for a SNIa survey assuming different magnitude cuts ({\em bottom panels}). {\em Left}: as a function of distance to the source, i.e. $r(z)$, with the source fixed at $\bar{r}(z=2)$; {\em Right}: as a function of $\bar{r}(z)$ (i.e. source distance) at a fixed $r=0.5\bar{r}$, thus examining the halfway point between source and observer. Dashed lines indicate negative values. We only show the amplitudes assuming the biases are constructed from a Power Law + Peak chirp mass distribution for GWs, as the difference with the Broken Power Law model is negligible. Similarly, for SNIa we only show the biases for a population of SNIa of fixed intrinsic luminosity, as the resulting amplitudes differ only slightly. Note that in the bottom left plot the blue line stops at lower redshifts due to low number of sources. We also plot in grey (for illustration purposes only in the bottom plot) the corresponding amplitudes when setting both biases to zero (solid), only magnification (dotted), and only evolution bias (dashdotted).
    }
    \label{fig:lensing_}
\end{figure}

We then focus on the lensing amplitude, addressing the same set of biases as in figure \ref{fig:doppler}. On the left of figure \ref{fig:lensing_} we fix the source at $z=2$ and explore $A_L$ as a function of distance $r(z)$ to the object, thus looking at the integrand on the right-hand side of eq. \eqref{eq:coefficients}. As we previously suggested in \cite{Us}, the shape of this curve is strictly dependent on the value of the biases, together with the zero-crossing separating magnification near the source from de-magnification away from it. 
This is clear from the bottom left panel of figure \ref{fig:lensing_} showing the lensing amplitude for SNIa surveys with different limiting apparent magnitudes: \refrep{higher} values of $m_{lim}$ (\refrep{lower} threshold), thus \refrep{lower} magnification bias, shift the zero crossing further from the observer, shrinking the redshift range in which magnification occurs (i.e. $A_L >0$). Instead, for an ET-like experiment the difference in the lensing amplitudes calculated using different SNR thresholds is close to negligible, given the small values of $s$ shown in figure \ref{fig:mag_biases_3G}.

On the right-hand side we plot the lensing amplitude at the halfway distance between observer and source, i.e. $r=0.5\bar{r}$, as a function of source distance $\bar{r}$.
Further, in the bottom plots we show in grey three additional curves: $A_L$ with both biases set to zero (solid line), only magnification bias (dotted) and only evolution bias (dashdotted). We plot these only in the bottom panels for SNIa as for GWs the difference with or without biases was minimal. These clearly show the strong dependence of the lensing amplitude on the magnification bias, whilst the evolution bias, despite being present in the expression in eq. \eqref{eq:AL}, has a negligible contribution.

\subsection{Impact on the angular power spectrum}

After examining the impact of the biases on the amplitudes of certain corrections to the number count, we explore their relevance in the angular power spectrum. Using a modified version of the code \texttt{CAMB}\footnote{https://github.com/cmbant/CAMB} we produced in earlier work \cite{Us}, we can compute angular power spectra supplying different values of both magnification and evolution bias. Whilst for GWs we use broad Gaussian windows ($\sigma=0.2$) for the sample bins, motivated by the large uncertainty in the estimation of the luminosity distance of GWs \cite{GWTC2,GWTC3}, for SNIa we chose smaller bins ($\sigma=0.1$), as the related distance uncertainties for LSST will be smaller \cite{Pantheon,Howlett_2017,Brout_2022}.

\begin{figure}
    \centering
    \includegraphics[width=0.9\textwidth]{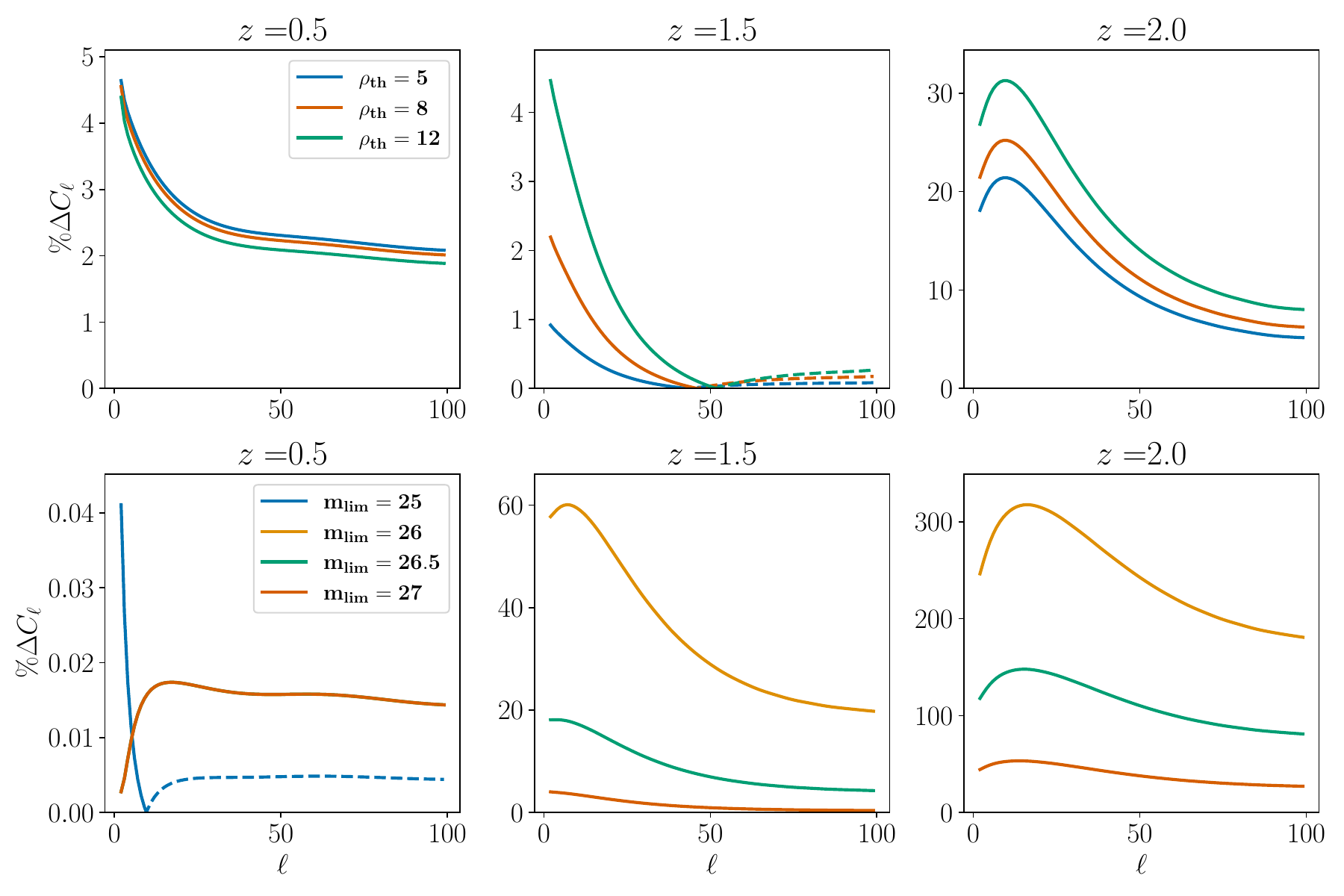}
    \caption{Impact of the models of the biases considered on auto-correlation angular power spectra for GWs as seen by an ET-like experiment ({\em top}) and for a magnitude limited SNIa survey ({\em bottom}). We plot the percentage difference between angular power spectra with biases as opposed to without them. The bottom left panel shows models with limiting magnitude $m_{lim}>25$ overlapping; bottom right is missing $m_{lim}=25$ as this is outside our definition of validity limit for the bias for SNIa (i.e. less than $2\sigma$ of the magnitude distribution is observed). Dashed lines represent negative values.}
    \label{fig:auto_cells}
\end{figure}

Hence, we plot the percentage difference in the angular power spectrum at each scale when accounting for the biases compared to the case where they are both set to zero for both GWs and SNIa, respectively top and bottom of figure \ref{fig:auto_cells}. We do this for three different redshifts, noting that $z=1.5$ is outside what we defined as the validity range 
of the bias for SNIa assuming a limiting apparent magnitude $m_{lim}=25$, and thus this particular case is not shown in the bottom right panel of figure \ref{fig:auto_cells}. We recall that the SNIa validity limit was fixed at the distance at which $2\sigma$ of the magnitude distribution is observed.

\refrep{One can immediately note the difference between the two tracers. Whilst SNIa show an increasing strength of the biases at higher redshifts, the models considered here for GWs impact differently. In fact, at low redshift the effect of the biases is simply at a percentage level, approaching a negligible one around $z=1.5$, and rising steeply after. This follows directly from the models of the biases studied here: the evolution bias (see figure \ref{fig:evo_bias_3G}) is close to $-2$ at very low redshifts, crosses $0$ around $z=1.5$, and then rises further.}

\begin{figure}
    \centering
    \includegraphics[width=0.9\textwidth]{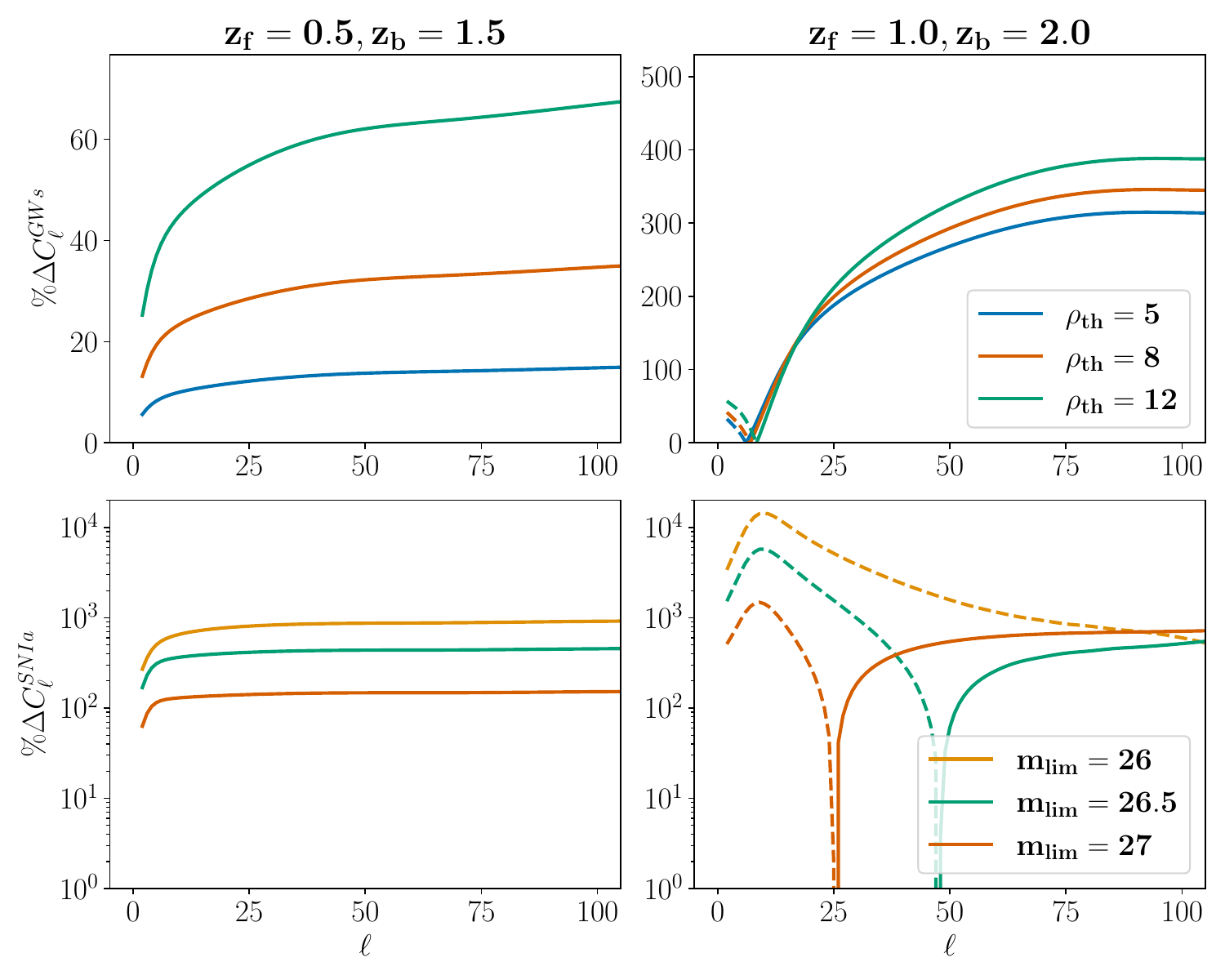}
    \caption{Impact of the models of the biases considered on cross-bins correlations, plotting the percentage difference between angular power spectra with biases and those without. {\em Left}: correlating a background bin at $z_b=1.5$ with a foreground one at $z_f=0.5$; {\em right}: the same with background one set to $z_b=2$ and foreground one to $z_f=1$. As in figure \ref{fig:auto_cells}, {\em top} is for GWs for an ET-like experiment, and {\em bottom} for SNIa. The bottom plots lack the model with limiting magnitude $m_{lim}=25$ as it is outside the validity limit.}
    \label{fig:cross_bins}
\end{figure}

We then examine the impact of the models of the biases considered when cross-correlating different redshift bins. We show two different examples of this: setting a background tracer at $z_b=1.5$ and a foreground one at $z_f=0.5$, and similarly with $z_b=2.0$ and $z_f=1.0$. This is shown in figure \ref{fig:cross_bins}, with GWs in the top panels and SNIa in the bottom ones. Similarly to figure \ref{fig:auto_cells}, the plots report the percentage difference between the angular power spectra with biases compared to those with the biases set to zero. It is clear how not accounting for the magnification and evolution biases can lead to large differences, depending on the tracer considered. On the top left panel, cross-bins angular power spectra for GWs show percentage level difference, although the contrast increases drastically by roughly an order of magnitude when shifting to higher redshifts (on the top right). On the other hand, angular power spectra built using SNIa already show differences of \refrep{several times} those with biases set to zero; going to higher redshifts can push the percentage difference to \refrep{a few order of magnitude}. In all cases, the difference \refrep{tends} to a constant value at a larger value of $\ell$ which depends on the redshift considered.

The large differences clearly show that cross-correlations require considering the impact of the magnification and evolution biases for both types of transient tracers. Additionally, cross-bin correlations allow for the different bias models to be distinguishable even at lower redshifts. This is clear from the plots on the left-hand side in figure \ref{fig:cross_bins}, where each model produces a distinct difference with respect to the angular power spectrum without biases. Such contrast is not seen in the auto-correlations with GWs in the top panels of figure \ref{fig:auto_cells} even at $z=1.5$, where the two models are both roughly of the same order of magnitude. SNIa models already showed substantial differences in auto-correlations at high redshift, as shown in the bottom right panel of figure \ref{fig:auto_cells}. However, the distinction becomes more significant in the cross-bins correlation, with differences of even an order of magnitude arise between different limiting magnitude models, as shown in the bottom plots of figure \ref{fig:cross_bins}.

\begin{figure}
    \centering
    \includegraphics[width=0.9\textwidth]{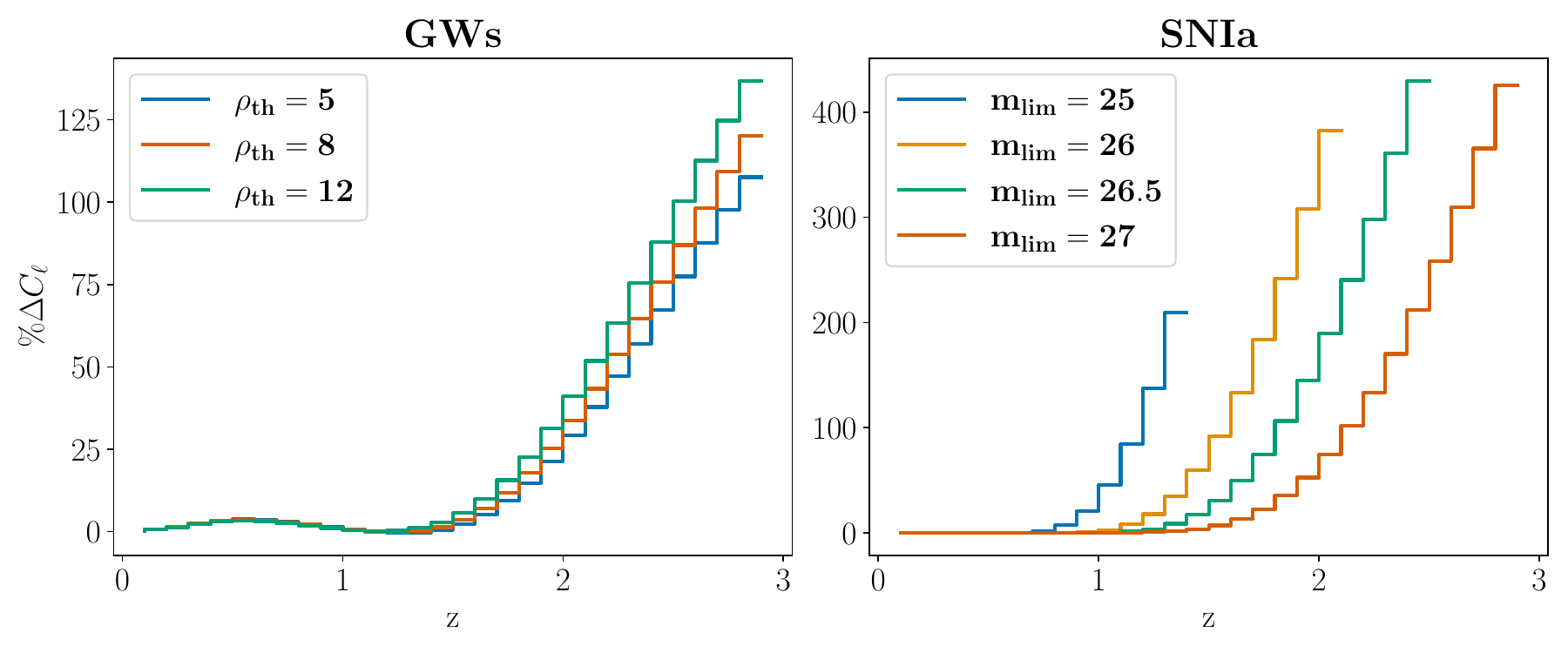}
    \caption{Impact of the biases considered on the angular power spectrum at $\ell=10$ across redshift. As before, we plot the percentage difference between the $C_{\ell}$ with biases as opposed to those without. }
    \label{fig:same_ell}
\end{figure}

Finally, we explore the impact of the biases modelled in this paper on the auto-correlation angular power spectrum across redshift. We plot the percentage difference between the non-zero biases case with 
respect to the zero bias case 
as a function of redshift in figure \ref{fig:same_ell}. The step in the plot represents the binning used for both GWs and SNIa, respectively $z=0.2$ and $z=1$; thus, we decided to avoid a smooth curve for clarity. On the LHS we plot the difference when calculated for GWs observed with an ET-like detector, and on the RHS the same for SNIa from a magnitude-limited survey. 
The former shows (positive) deviation from zero from around $z=1$, growing almost linearly with redshift. It also shows that before such a point, the angular power spectrum with biases accounted for \refrep{grows slightly and turns back towards zero}. An explanation of this could be traced to the behaviour of the magnification and evolution biases for ET and CE (shown in figure \ref{fig:evo_bias_3G}). Since $s^{GW}$ is significantly close to zero, the largest contribution will be given by the evolution bias; and considering that $b_e^{GW}$ \refrep{crosses zero around} $z<1.5$, this might explain the dip in the percentage difference for GWs in figure \ref{fig:same_ell}.

On the other hand, as shown in figure \ref{fig:biases_SNIa}, the evolution bias of SNIa was very small, whereas $s^{SNIa}$ was substantially larger than the GWs case. The Doppler kernel in figure \ref{fig:doppler} and the lensing amplitude in figure \ref{fig:lensing_}, the correction terms for SNIa were dominated by the impact of the large magnification bias; similarly, the difference in the auto-correlation angular power spectra between the case with biases and the one with them set to zero, is significantly impacted by the magnification bias. This is clear by the shape of the curves shown in RHS of figure \ref{fig:same_ell}: the difference is zero until the redshift value at which the corresponding $s^{SNIa}$ starts growing. 

Figure \ref{fig:same_ell} further stresses that for SNIa the models can be clearly distinguished from $z\sim1$; however, whereas auto-correlation angular power spectra with GWs do highlight the need for accounting for the biases, they do not provide large enough differences between the models considered. Such distinction is only seen when cross-correlating separate redshift bins, as seen in figure \ref{fig:cross_bins}.

\section{Summary and conclusions}\label{sec:conclusions}
By modelling and ultimately measuring the bias properties of tracers, we can use them as probes of large-scale structure and cosmology. The bias properties of galaxies is a well-studied topic; with the advent of large data sets of transient sources, we need to learn how to describe their biases equally well. As we have eluded to, there are particular fundamental physics motivations for wanting to use tracers who most naturally `belong' in luminosity distance space, such as GW sources and SN1a.

The observed fluctuation of the source number counts, $\Delta_O$, is the basis of many clustering analyses \cite{Bonvin_2011, Challinor_2011}. When considering transient tracers such as GWs or SNIa, this has to be computed in luminosity distance space, as the two types of object do not carry direct information of their redshift \cite{Us}. The expression for the number counts contains several relativistic corrections, notably Doppler magnification and lensing. These depend on the magnification and evolution biases, which can significantly alter their amplitudes. 

In this paper, we first described the modelling of magnification and evolution biases of gravitational waves from binary black hole mergers. We initially carefully described the necessity of defining them with respect to a specific detector and not to any single ``intrinsic'' quantity, as opposed to the traditional case in galaxy clustering.
We also employed two different distributions of chirp masses, namely a Power Law + Peak and a Broken Power Law consistent with population analysis of the GWTC-3 catalogue by the LVK collaboration \cite{GWTC2,GWTC3}. Furthermore, we examined GW biases for three different SNR thresholds, noting that a higher $\rho_{th}$ results in larger values of the biases. The lower sensitivities of the present terrestrial detectors (relative to 3G expected sensitivity curves) yields a strongly negative evolution bias. This implies that sources are becoming sparser with redshift; however, given the expectation of a peak around cosmic dawn ($z\sim 2$), it is clear this is due to a decline in the detector's sensitivity at these redshifts. In fact, third-generation observatories ET and CE should instead be able to perfectly trace the evolution of GWs from BBH mergers. Consequently, their related magnification biases are particularly small up to high redshift. We note that the method described for GWs from BBH mergers can be equally applied to different sources of GWs, such as neutron stars mergers or black hole-neutron star binaries, provided the appropriate chirp mass distribution and merger rate are supplied.

Further, we explored the biases for SNIa, investigating the impact of changing the magnitude threshold. Whilst the magnification bias is especially sensitive to this, the evolution bias for the SNIa models considered is independent of it. Additionally, the former can reach much larger values than the ones seen for GWs, while the latter is close to flat around zero, implying a population density that remains close to constant across redshift.
We also explored the possibility of a population of supernovae of intrinsic luminosity evolving with redshift, using two different models proposed in the literature. We found that the magnification bias is only slightly altered, while the evolution one changes significantly depending on the model adopted (i.e. whether the intrinsic luminosity of SNIa decreases or increases with redshift).

Finally, we investigated the effects of these biases on relativistic corrections to the observed number counts and on the angular power spectrum. In both cases, the impact of the biases produces a significant difference with respect to cases where the biases were set to zero. We found that the lensing magnification is strongly dependent on the magnification bias, as expected. However, the relativistic Doppler correction shows a different behaviour depending on the tracer. For GWs it is sensitive to the evolution bias, whilst for SNIa the dominant bias is
the magnification bias. This is explained by the different values taken by both $s$ and $b_e$ for the two different tracers: the former being significantly small for GWs, the latter for SNIa. Therefore, with one of the two parameters closer to zero, the other has a stronger impact. 

The same effect occurs then when comparing angular power spectra with biases included to ones with biases set to zero, as in figure \ref{fig:same_ell}. This in turn would have an impact on any derived constraints from the angular power spectra in luminosity distance. For GWs, the difference between the angular power spectra is initially \refrep{small}, rising after the redshift for which $b_e^{GWs}>0$. For SNIa, the difference is null until the distance at which $s^{SNIa}>0$. 

Furthermore, when analysing the impact of the biases on the angular power spectrum, we found a percentage level difference in auto-correlations with GWs between spectra with biases accounted for and spectra with biases set to zero. When the same analysis is applied to SNIa, the differences are greatly increased at high redshifts, i.e. when $s^{SNIa}>0$. This shows the importance of accounting for the biases when computing angular power spectra, however, at least for GWs, the different models of the biases are distinguishable only at high redshifts. The separation between models is instead achieved more clearly when investigating cross-bin correlations between different redshifts. Figure \ref{fig:cross_bins} not only shows greatly increased differences with respect to the unbiased spectra, but also highlights the possibility of distinguishing the specific model of the bias. As these parameters are strictly dependent on population properties, distinguishing between them could help us infer details of the mass distribution of these tracers. 

Having set up frameworks with which to model transient biases, our next step will be to understand how these impact constraints from cross-correlations in the  era of stage IV galaxy surveys, and the 3G era of gravitational wave detection. This impacts not only 3G cosmology, but may also be highly relevant for astrophysics and compact object formation channels, if we find that population properties can be simultaneously constrained.

\acknowledgments
We are pleased to thank Michelle Lochner, Michele Mancarella, and Roy Maartens (and others) for useful discussions. S.Z. acknowledges support from the Perren Fund, University of London. JF thanks the support of Funda\c{c}\~{a}o para a Ci\^{e}ncia e a Tecnologia (FCT) through the
research grants UIDB/04434/2020 and UIDP/04434/2020 and through
the Investigador FCT Contract No. 2020.02633.CEECIND/CP1631/CT0002.
JF also thanks the hospitality of the Astronomy Unit of QMUL and the University of the Western Cape where part of this work was developed. We thank the support of FCT and the Portuguese Association of Researcher and Students in the UK (PARSUK) under the Portugal - United Kingdom exchange program Bilateral Research Fund (BRF). T.B. is supported by ERC Starting Grant \textit{SHADE} (grant no.~StG 949572) and a Royal Society University Research Fellowship (grant no.~URF$\backslash$R1$\backslash$180009). CC is supported by the UK Science \& Technology Facilities Council Consolidated Grant ST/T000341/1. 

\appendix

\section{BBH mass distributions}\label{sec:mass_distributions}
The expressions derived for the magnification and evolution bias both require an appropriate distribution of chirp masses, $\phi(\M)$. A population analysis of the second LIGO-Virgo Gravitational Waves Transient Catalog (GWTC-2) produced distributions of masses of the primary compact object, together with a conditional distribution for the ratio $q={m_{2}}/{m_{1}}$.
The models favoured are the following: 
\begin{itemize}
    \item \textit{Power Law + Peak}: 
    
    Motivated by the idea that the mass loss undergone by pulsational pair-instability supernovae could lead to a pileup of BBH mergers before the pair-instability gap. The distribution is given by:
    \bea \label{eq:pl+p}
        \pi(m_{1}|\lambda_{peak},\alpha,m_{min},\delta_{m},m_{max},\mu_{m},\sigma_{m}) =
        [(1-\lambda_{peak})\mathfrak{B}(m_{1}|-\alpha,m_{max})\nn\\
        +\lambda_{peak}G(m_{1}|\mu_{m},\sigma_{m})]S(m_{1}|m_{min},\delta_{m})
    \eea
    
    where $\mathfrak{B}$ is a normalised power law of spectral index $\alpha$ and mass cut-off $m_{max}$, $G(m_{1}|\mu_{m},\sigma_{m})$ is a normalised gaussian with mean $\mu$ and width $\sigma$. Further, $S(m_{1}|m_{min})$ is a smoothing function defined as:
    \be \label{eq:smoothing}
        S(m_{1}|m_{min}) = 
        \begin{cases}
            0 ,& \text{if } m < m_{min}\\
            [f(m-m_{min},\delta_{m})+1]^{-1} ,& \text{if } m_{min} \le m < m_{min} +\delta_{m}\\
            1 ,& \text{if } m \ge m_{min} + \delta_{m}\\
        \end{cases}
    \ee
    with
    \be
        f(m',\delta_{m}) = \exp\left(\frac{\delta_{m}}{m'} + \frac{\delta_{m}}{m'-\delta_{m}}\right)
    \ee
    
    \item \textit{Broken Power Law}:
    
    This model is an extension to a simple truncated power law one, motivated by the potential tapering of the primary mass distribution at high masses. It also uses a smoothing function $S(m_{1}|m_{min})$ as in equation \eqref{eq:smoothing}. 
    The model is the following:
    \be \label{eq:bpl}
    \pi(m_{1}|\alpha_{1},\alpha_{2},m_{min},m_{max}) \propto
    \begin{cases}
        m_{1}^{\alpha_{1}}S(m_{1}|m_{min},\delta_{m}) ,& \text{if } m_{min} < m_{1} < m_{break}\\
        m_{1}^{\alpha_{2}}S(m_{1}|m_{min},\delta_{m}) ,& \text{if } m_{break} < m_{1} < m_{max}\\
        0 ,& \text{otherwise}\\
    \end{cases}
    \ee
    where $m_{break}$ is defined by
    \be
    m_{break} = m_{min} + b(m_{max}-m_{min})
    \ee
    with $b$ being the fraction of the way between $m_{min}$ and $m_{max}$ at which the primary mass distribution undergoes a break.
\end{itemize}

Both models see the same conditional mass ratio distribution form, albeit the relevant parameters have different values:
\be \label{eq:conditional}
\pi(q|\beta,m_{1},m_{min},\delta_{m}) \propto q^{\beta_{q}}S(qm_{1}|m_{min},\delta_{m}).
\ee

\section{Chirp Mass distribution}\label{sec:chirp_calc}

We want to compute the PDF $h(z)$ of the chirp mass $\mathcal{M}$ given the distributions of primary and secondary masses, $g(m_1)$ and $f(m_2)$. Let us recast these random variables (for simplicity) in $m_2 = x$,$m_1 = y$, $\mathcal{M} = z$, and their relation is given by the chirp mass definition:
\bea \label{eq:chirp_xy}
z = \frac{(xy)^{3/5}}{(x+y)^{1/5}} \, .\
\eea

Now, consider the following transformation:
\bea
\begin{cases}\label{eq:trasformation}
    z = z(x,y) \, \\
    u = y \, 
\end{cases}
\eea
The joint PDF $p(z,u)$ of the new variables $z,u$ is related to that of $x,y$ by the conservation of the volume in the space of probability
\bea \label{eq:multiplication}
f(x)g(y)\d x\d y = p(z,u)\d u\d z\, ,
\eea
where we have assumed that $f(x),g(y)$ are independent. The differentials are related by the determinant of the Jacobian matrix of the transformation:
\bea \label{eq:dudz}
\d x\d y = |J(z,u)|\d u\d z\, .
\eea
Taking $x_i$ as a root of the equation $z=z(x,y)$, and the fact that, from \eqref{eq:trasformation}, $y(z,u)=u$, we can write the Jacobian as:
\bea\label{eq:Jacobian}
|J(z,u)| = \frac{\partial x_i(z,u)}{\partial z}\, .\
\eea
Substituting \eqref{eq:Jacobian} and \eqref{eq:dudz} into \eqref{eq:multiplication}, and summing over \refrep{the possible roots}, we find the combined PDF
\bea
p(z,u) = \sum_{x_i} f(x_i(z,u))g(u)\frac{\partial x_i(z,u)}{\partial z} \, ,
\eea
and to obtain the PDF for $z$, $h(z)$, we marginalise over $u$:
\bea \label{eq:distribution}
h(z) = \sum_{x_i}\int \d u\ f(x_i(z,u))g(u)\frac{\partial x_i(z,u)}{\partial z} \, .
\eea
Now, looking at \eqref{eq:chirp_xy}, we need to find the solutions for $x(z,u)$. We can recast the equation as a third order equation for x of the form
\bea
x^3+bx+c=0\, ,\quad \text{with} \quad b=-\frac{z^5}{u^3}, \quad c=-\frac{z^5}{u^2} \,.
\eea
Depending on the sign of the delta of the cubic equation above
\bea
\Delta = \frac{c^2}{4}+\frac{b^3}{27}=\frac{z^{10}}{u^4}\left(\frac1{4}-\frac{z^5}{27u^5}\right) \, ,
\eea
we have three separate cases: $\Delta < 0$, $\Delta = 0$, $\Delta > 0$.
For $\Delta = 0$, the three (real) solutions are valid at only point in the domain, corresponding to $u = (4/27)^{1/5}z$. Therefore, since we are interested in the distribution \eqref{eq:distribution} that involves an integral over $u$, we can simply neglect them. We are then left with $\Delta < 0$ which results in $u < (4/27)^{1/5}z$, and $\Delta > 0$, which gives the range $u < (4/27)^{1/5}z$.

However, recalling that $z=\M$ and $u=y=m_1$, then the first integral is given when $m_1>(4/27)^{1/5}\M$. Notably, this includes the regime of $m_1 > m_2$, which results in $m_1 > 2^{1/5}\M$. Conversely, the second integral describes the regime for which $m_1 < m_2$. However, we define the primary and secondary masses as, respectively, $m_1$ and $m_2$, and we also make the assumption that $m_1 \geq m_2$ always. Therefore, we can safely neglect the second integral. The functions $f(x_1)$ and $g(u)$ corresponds then to, respectively, the conditional PDF of the secondary mass and the PDF of the primary one:

\bea \label{eq:chirp_result}
h(z) &=& \int_{u>(4/27)^{1/5}z} \d u f(x_{1}(z,u))g(u)\frac{\dd{x_{1}(z,u)}}{\dd{z}} \,.
\eea

\begin{figure}
    \centering
    \includegraphics[width=0.9\textwidth]{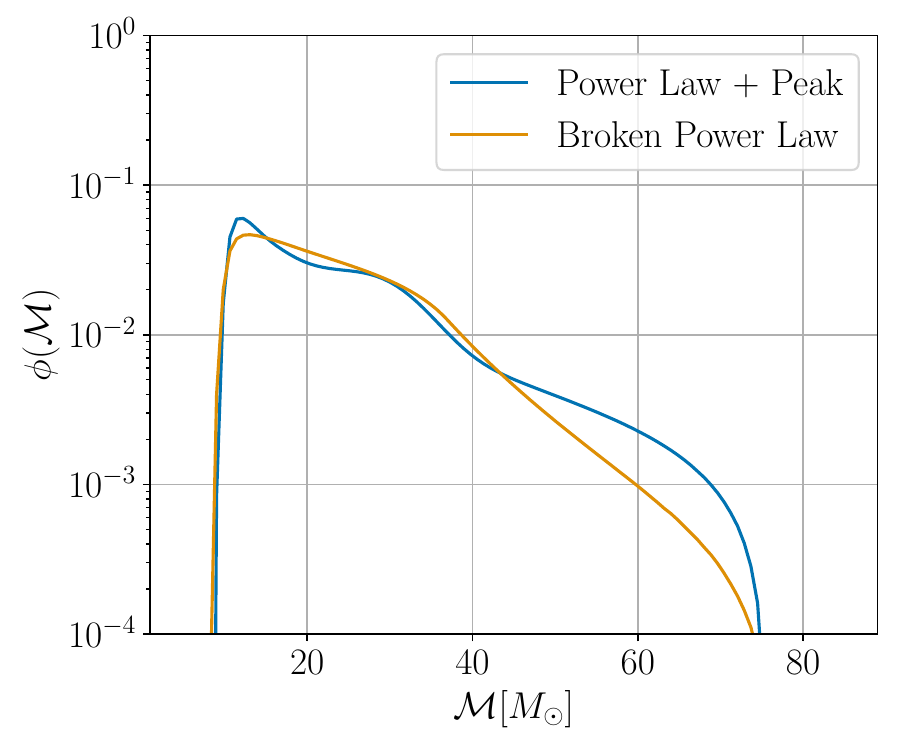}
    \caption{Comparison of the chirp mass pdf for two types of primary mass distribution.}
    \label{fig:chirps_plot}
\end{figure}

We can then apply the distributions discussed in appendix \ref{sec:mass_distributions} to calculate the chirp mass pdf for a Power Law + Peak and for a Broken Power Law primary distribution. We plot the result in figure \ref{fig:chirps_plot}. The two distributions are significantly similar to each other, with the largest difference at higher masses.

\vfill

\bibliographystyle{JHEP}
\bibliography{biasbib}

\providecommand{\href}[2]{#2}\begingroup\raggedright\begin{thebibliography}{10}

\bibitem{Martinez_1999}
H.J.~Martinez, M.E.~Merchan, C.A.~Valotto and D.G.~Lambas, \emph{Quasar-galaxy and {AGN}-galaxy cross-correlations}, \href{https://doi.org/10.1086/306973}{\emph{The Astrophysical Journal} {\bfseries 514} (1999) 558}.

\bibitem{Padmanabhan_2007}
N.~Padmanabhan, D.J.~Schlegel, U.~Seljak, A.~Makarov, N.A.~Bahcall, M.R.~Blanton et~al., \emph{The clustering of luminous red galaxies in the sloan digital sky survey imaging data}, \href{https://doi.org/10.1111/j.1365-2966.2007.11593.x}{\emph{Monthly Notices of the Royal Astronomical Society} {\bfseries 378} (2007) 852}.

\bibitem{Bonvin_2008}
C.~Bonvin, \emph{Effect of peculiar motion in weak lensing}, \href{https://doi.org/10.1103/physrevd.78.123530}{\emph{Physical Review D} {\bfseries 78} (2008) }.

\bibitem{Hui_2008}
L.~Hui, E.~Gazta{\~{n} }aga and M.~LoVerde, \emph{Anisotropic magnification distortion of the 3d galaxy correlation. {II}. fourier and redshift space}, \href{https://doi.org/10.1103/physrevd.77.063526}{\emph{Physical Review D} {\bfseries 77} (2008) }.

\bibitem{Chen_2008}
J.~Chen, \emph{The galaxy cross-correlation function as a probe of the spatial distribution of galactic satellites}, \href{https://doi.org/10.1051/0004-6361:20079184}{\emph{A\&A} {\bfseries 494} (2008) 867}.

\bibitem{Yoo_2009}
J.~Yoo, A.L.~Fitzpatrick and M.~Zaldarriaga, \emph{New perspective on galaxy clustering as a cosmological probe: General relativistic effects}, \href{https://doi.org/10.1103/physrevd.80.083514}{\emph{Physical Review D} {\bfseries 80} (2009) }.

\bibitem{Bonvin_2011}
C.~Bonvin and R.~Durrer, \emph{What galaxy surveys really measure}, \href{https://doi.org/10.1103/physrevd.84.063505}{\emph{Physical Review D} {\bfseries 84} (2011) }.

\bibitem{Challinor_2011}
A.~Challinor and A.~Lewis, \emph{Linear power spectrum of observed source number counts}, \href{https://doi.org/10.1103/physrevd.84.043516}{\emph{Physical Review D} {\bfseries 84} (2011) }.

\bibitem{Jeong_2012}
D.~Jeong, F.~Schmidt and C.M.~Hirata, \emph{Large-scale clustering of galaxies in general relativity}, \href{https://doi.org/10.1103/physrevd.85.023504}{\emph{Physical Review D} {\bfseries 85} (2012) }.

\bibitem{Tansella_2018}
V.~Tansella, C.~Bonvin, R.~Durrer, B.~Ghosh and E.~Sellentin, \emph{The full-sky relativistic correlation function and power spectrum of galaxy number counts. part i: theoretical aspects}, \href{https://doi.org/10.1088/1475-7516/2018/03/019}{\emph{Journal of Cosmology and Astroparticle Physics} {\bfseries 2018} (2018) 019}.

\bibitem{Castorina_2022}
E.~Castorina and E.D.~Dio, \emph{The observed galaxy power spectrum in general relativity}, \href{https://doi.org/10.1088/1475-7516/2022/01/061}{\emph{Journal of Cosmology and Astroparticle Physics} {\bfseries 2022} (2022) 061}.

\bibitem{Kaiser_1992}
N.~Kaiser, \emph{Weak gravitational lensing of distant galaxies}, \href{https://doi.org/10.1086/171151}{\emph{The Astrophysical Journal} {\bfseries 388} (1992) 272}.

\bibitem{Hamilton_1998}
A.J.S.~Hamilton, \emph{Linear redshift distortions: A review},  in \emph{Astrophysics and Space Science Library}, pp.~185--275, Springer Netherlands (1998), \href{https://doi.org/10.1007/978-94-011-4960-0_17}{DOI}.

\bibitem{Sloan}
J.M.~Kubo, J.~Annis, F.M.~Hardin, D.~Kubik, K.~Lawhorn, H.~Lin et~al., \emph{{THE} {SLOAN} {NEARBY} {CLUSTER} {WEAK} {LENSING} {SURVEY}}, \href{https://doi.org/10.1088/0004-637x/702/2/l110}{\emph{The Astrophysical Journal} {\bfseries 702} (2009) L110}.

\bibitem{Jain_2011}
B.~Jain and M.~Lima, \emph{{Magnification effects on source counts and fluxes}}, \href{https://doi.org/10.1111/j.1365-2966.2010.17505.x}{\emph{Monthly Notices of the Royal Astronomical Society} {\bfseries 411} (2011) 2113}.

\bibitem{Mason_2015}
C.A.~Mason, T.~Treu, K.B.~Schmidt, T.E.~Collett, M.~Trenti, P.J.~Marshall et~al., \emph{Correcting the $z\sim8$ galaxy luminosity function for gravitational lensing magnification bias}, \href{https://doi.org/10.1088/0004-637x/805/1/79}{\emph{The Astrophysical Journal} {\bfseries 805} (2015) 79}.

\bibitem{Leung_2018}
E.~Leung, T.~Broadhurst, J.~Lim, J.M.~Diego, T.~Chiueh, H.-Y.~Schive et~al., \emph{Magnification bias of distant galaxies in the hubble frontier fields: Testing wave versus particle dark matter predictions}, \href{https://doi.org/10.3847/1538-4357/aacdad}{\emph{The Astrophysical Journal} {\bfseries 862} (2018) 156}.

\bibitem{Duncan_2022}
C.A.J.~Duncan, J.~Harnois-D{\'{e} }raps, L.~Miller and A.~Langedijk, \emph{On cosmological bias due to the magnification of shear and position samples in modern weak lensing analyses}, \href{https://doi.org/10.1093/mnras/stac1809}{\emph{Monthly Notices of the Royal Astronomical Society} {\bfseries 515} (2022) 1130}.

\bibitem{Libanore_2022}
S.~Libanore, M.~Artale, D.~Karagiannis, M.~Liguori, N.~Bartolo, Y.~Bouffanais et~al., \emph{Clustering of gravitational wave and supernovae events: a multitracer analysis in luminosity distance space}, \href{https://doi.org/10.1088/1475-7516/2022/02/003}{\emph{Journal of Cosmology and Astroparticle Physics} {\bfseries 2022} (2022) 003}.

\bibitem{LSST_2021}
B.~Abolfathi, D.~Alonso, R.~Armstrong, {\'{E}}.~Aubourg, H.~Awan, Y.N.~Babuji et~al., \emph{The {LSST} {DESC} {DC}2 simulated sky survey}, \href{https://doi.org/10.3847/1538-4365/abd62c}{\emph{The Astrophysical Journal Supplement Series} {\bfseries 253} (2021) 31}.

\bibitem{Sanchez_2022}
B.O.~S{\'{a} }nchez, R.~Kessler, D.~Scolnic, R.~Armstrong, R.~Biswas, J.~Bogart et~al., \emph{{SNIa} cosmology analysis results from simulated {LSST} images: From difference imaging to constraints on dark energy}, \href{https://doi.org/10.3847/1538-4357/ac7a37}{\emph{The Astrophysical Journal} {\bfseries 934} (2022) 96}.

\bibitem{LSST_book}
L.S.~Collaboration, P.A.~Abell, J.~Allison, S.F.~Anderson, J.R.~Andrew, J.R.P.~Angel et~al., \emph{Lsst science book, version 2.0},  2009.

\bibitem{Libanore_2021}
S.~Libanore, M.C.~Artale, D.~Karagiannis, M.~Liguori, N.~Bartolo, Y.~Bouffanais et~al., \emph{Gravitational wave mergers as tracers of large scale structures}, \href{https://doi.org/10.1088/1475-7516/2021/02/035}{\emph{Journal of Cosmology and Astroparticle Physics} {\bfseries 2021} (2021) 035}.

\bibitem{Scelfo_2018}
G.~Scelfo, N.~Bellomo, A.~Raccanelli, S.~Matarrese and L.~Verde, \emph{{GW}{\texttimes}{LSS}: chasing the progenitors of merging binary black holes}, \href{https://doi.org/10.1088/1475-7516/2018/09/039}{\emph{Journal of Cosmology and Astroparticle Physics} {\bfseries 2018} (2018) 039}.

\bibitem{Scelfo_2020}
G.~Scelfo, L.~Boco, A.~Lapi and M.~Viel, \emph{Exploring galaxies-gravitational waves cross-correlations as an astrophysical probe}, \href{https://doi.org/10.1088/1475-7516/2020/10/045}{\emph{Journal of Cosmology and Astroparticle Physics} {\bfseries 2020} (2020) 045}.

\bibitem{Scelfo_2022}
G.~Scelfo, M.~Spinelli, A.~Raccanelli, L.~Boco, A.~Lapi and M.~Viel, \emph{Gravitational waves {\texttimes} {HI} intensity mapping: cosmological and astrophysical applications}, \href{https://doi.org/10.1088/1475-7516/2022/01/004}{\emph{Journal of Cosmology and Astroparticle Physics} {\bfseries 2022} (2022) 004}.

\bibitem{Scelfo_2022_2}
G.~Scelfo, M.~Berti, A.~Silvestri and M.~Viel, \emph{Testing gravity with gravitational waves {\texttimes} electromagnetic probes cross-correlations}, \href{https://doi.org/10.1088/1475-7516/2023/02/010}{\emph{Journal of Cosmology and Astroparticle Physics} {\bfseries 2023} (2023) 010}.

\bibitem{Sathyaprakash_2010}
B.S.~Sathyaprakash, B.F.~Schutz and C.V.D.~Broeck, \emph{Cosmography with the einstein telescope}, \href{https://doi.org/10.1088/0264-9381/27/21/215006}{\emph{Classical and Quantum Gravity} {\bfseries 27} (2010) 215006}.

\bibitem{Sathyaprakash_2012}
B.~Sathyaprakash, M.~Abernathy, F.~Acernese, P.~Ajith, B.~Allen, P.~Amaro-Seoane et~al., \emph{Scientific objectives of einstein telescope}, \href{https://doi.org/10.1088/0264-9381/29/12/124013}{\emph{Classical and Quantum Gravity} {\bfseries 29} (2012) 124013}.

\bibitem{Maggiore_2020}
M.~Maggiore, C.V.D.~Broeck, N.~Bartolo, E.~Belgacem, D.~Bertacca, M.A.~Bizouard et~al., \emph{Science case for the einstein telescope}, \href{https://doi.org/10.1088/1475-7516/2020/03/050}{\emph{Journal of Cosmology and Astroparticle Physics} {\bfseries 2020} (2020) 050}.

\bibitem{Punturo_2010}
M.~Punturo, M.~Abernathy, F.~Acernese, B.~Allen, N.~Andersson, K.~Arun et~al., \emph{The third generation of gravitational wave observatories and their science reach}, \href{https://doi.org/10.1088/0264-9381/27/8/084007}{\emph{Classical and Quantum Gravity} {\bfseries 27} (2010) 084007}.

\bibitem{Evans_2021}
M.~Evans, R.X.~Adhikari, C.~Afle, S.W.~Ballmer, S.~Biscoveanu, S.~Borhanian et~al., \emph{A horizon study for cosmic explorer: Science, observatories, and community},  2021.

\bibitem{Reitze_2019}
D.~Reitze, R.X.~Adhikari, S.~Ballmer, B.~Barish, L.~Barsotti, G.~Billingsley et~al., \emph{Cosmic explorer: The u.s. contribution to gravitational-wave astronomy beyond ligo},  2019.

\bibitem{Mpetha_2022}
C.T.~Mpetha, G.~Congedo and A.~Taylor, \emph{Future prospects on testing extensions to $\lambda$cdm through the weak lensing of gravitational waves}, \href{https://doi.org/10.1103/physrevd.107.103518}{\emph{Physical Review D} {\bfseries 107} (2023) }.

\bibitem{Peron_2023}
M.~Peron, S.~Libanore, A.~Ravenni, M.~Liguori and M.C.~Artale, \emph{Clustering of binary black hole mergers: a detailed analysis of the eagle+mobse simulation},  2023.

\bibitem{Rigby_2011}
E.E.~Rigby, P.N.~Best, M.H.~Brookes, J.A.~Peacock, J.S.~Dunlop, H.J.A.~Röttgering et~al., \emph{The luminosity-dependent high-redshift turnover in the steep spectrum radio luminosity function: clear evidence for downsizing in the radio-{AGN} population}, \href{https://doi.org/10.1111/j.1365-2966.2011.19167.x}{\emph{Monthly Notices of the Royal Astronomical Society} {\bfseries 416} (2011) 1900}.

\bibitem{Us}
J.~Fonseca, S.~Zazzera, T.~Baker and C.~Clarkson, \emph{The observed number counts in luminosity distance space},  2023.

\bibitem{Namikawa}
T.~Namikawa, \emph{Analyzing clustering of astrophysical gravitational-wave sources: luminosity-distance space distortions}, \href{https://doi.org/10.1088/1475-7516/2021/01/036}{\emph{Journal of Cosmology and Astroparticle Physics} {\bfseries 2021} (2021) 036}.

\bibitem{Maartens_2021}
R.~Maartens, J.~Fonseca, S.~Camera, S.~Jolicoeur, J.-A.~Viljoen and C.~Clarkson, \emph{Magnification and evolution biases in large-scale structure surveys}, \href{https://doi.org/10.1088/1475-7516/2021/12/009}{\emph{Journal of Cosmology and Astroparticle Physics} {\bfseries 2021} (2021) 009}.

\bibitem{Mastrogiovanni_2023}
S.~Mastrogiovanni, C.~Bonvin, G.~Cusin and S.~Foffa, \emph{Detection and estimation of the cosmic dipole with the einstein telescope and cosmic explorer}, \href{https://doi.org/10.1093/mnras/stad430}{\emph{Monthly Notices of the Royal Astronomical Society} {\bfseries 521} (2023) 984}.

\bibitem{LIGOScientific:2021djp}
{\scshape LIGO Scientific, VIRGO, KAGRA} collaboration, \emph{Gwtc-3: Compact binary coalescences observed by ligo and virgo during the second part of the third observing run},  \href{https://arxiv.org/abs/2111.03606}{{\ttfamily 2111.03606}}.

\bibitem{GWTC3}
T.L.S.~Collaboration, the Virgo~Collaboration, the KAGRA~Collaboration, R.~Abbott, T.D.~Abbott, F.~Acernese et~al., \emph{The population of merging compact binaries inferred using gravitational waves through gwtc-3},  \href{https://arxiv.org/abs/2111.03634}{{\ttfamily 2111.03634}}.

\bibitem{GWTC2}
R.~Abbott, T.D.~Abbott, S.~Abraham, F.~Acernese, K.~Ackley, A.~Adams et~al., \emph{Population properties of compact objects from the second {LIGO}{\textendash}virgo gravitational-wave transient catalog}, \href{https://doi.org/10.3847/2041-8213/abe949}{\emph{The Astrophysical Journal Letters} {\bfseries 913} (2021) L7}.

\bibitem{Finn_1996}
L.S.~Finn, \emph{Binary inspiral, gravitational radiation, and cosmology}, \href{https://doi.org/10.1103/physrevd.53.2878}{\emph{Physical Review D} {\bfseries 53} (1996) 2878}.

\bibitem{Oguri_2018}
M.~Oguri, \emph{Effect of gravitational lensing on the distribution of gravitational waves from distant binary black hole mergers}, \href{https://doi.org/10.1093/mnras/sty2145}{\emph{Monthly Notices of the Royal Astronomical Society} {\bfseries 480} (2018) 3842}.

\bibitem{referee}
R.~Essick, \emph{Semianalytic sensitivity estimates for catalogs of gravitational-wave transients},  \href{https://arxiv.org/abs/2307.02765}{{\ttfamily 2307.02765}}.

\bibitem{YeFishback}
C.~Ye and M.~Fishbach, \emph{Cosmology with standard sirens at cosmic noon}, \href{https://doi.org/10.1103/physrevd.104.043507}{\emph{Physical Review D} {\bfseries 104} (2021) }.

\bibitem{MadauDickinson}
P.~Madau and M.~Dickinson, \emph{Cosmic star-formation history}, \href{https://doi.org/10.1146/annurev-astro-081811-125615}{\emph{Annual Review of Astronomy and Astrophysics} {\bfseries 52} (2014) 415}.

\bibitem{Oguri_2010}
M.~Oguri and P.J.~Marshall, \emph{Gravitationally lensed quasars and supernovae in future wide-field optical imaging surveys}, \href{https://doi.org/10.1111/j.1365-2966.2010.16639.x}{\emph{Monthly Notices of the Royal Astronomical Society} (2010) no}.

\bibitem{Oda_2005}
T.~Oda and T.~Totani, \emph{Deciphering the cosmic star formation history and the nature of type ia supernovae with future supernova surveys}, \href{https://doi.org/10.1086/431748}{\emph{The Astrophysical Journal} {\bfseries 630} (2005) 59}.

\bibitem{Yasuda_2010}
N.~Yasuda and M.~Fukugita, \emph{{LUMINOSITY} {FUNCTIONS} {OF} {TYPE} ia {SUPERNOVAE} {AND} {THEIR} {HOST} {GALAXIES} {FROM} {THE} {SLOAN} {DIGITAL} {SKY} {SURVEY}}, \href{https://doi.org/10.1088/0004-6256/139/1/39}{\emph{The Astronomical Journal} {\bfseries 139} (2009) 39}.

\bibitem{Hopkins_2006}
A.M.~Hopkins and J.F.~Beacom, \emph{On the normalization of the cosmic star formation history}, \href{https://doi.org/10.1086/506610}{\emph{The Astrophysical Journal} {\bfseries 651} (2006) 142}.

\bibitem{Madau_2014}
P.~Madau and M.~Dickinson, \emph{Cosmic star-formation history}, \href{https://doi.org/10.1146/annurev-astro-081811-125615}{\emph{Annual Review of Astronomy and Astrophysics} {\bfseries 52} (2014) 415}.

\bibitem{Totani_2008}
T.~Totani, T.~Morokuma, T.~Oda, M.~Doi and N.~Yasuda, \emph{Delay time distribution measurement of type ia supernovae by the subaru/{XMM}-newton deep survey and implications for the progenitor}, \href{https://doi.org/10.1093/pasj/60.6.1327}{\emph{Publications of the Astronomical Society of Japan} {\bfseries 60} (2008) 1327}.

\bibitem{Baldry_2003}
I.K.~Baldry and K.~Glazebrook, \emph{Constraints on a universal stellar initial mass function from ultraviolet to near-infrared galaxy luminosity densities}, \href{https://doi.org/10.1086/376502}{\emph{The Astrophysical Journal} {\bfseries 593} (2003) 258}.

\bibitem{Martinelli_2019}
M.~Martinelli and I.~Tutusaus, \emph{{CMB} tensions with low-redshift h0 and s8 measurements: Impact of a redshift-dependent type-ia supernovae intrinsic luminosity}, \href{https://doi.org/10.3390/sym11080986}{\emph{Symmetry} {\bfseries 11} (2019) 986}.

\bibitem{Linden_2009}
S.~Linden, J.-M.~Virey and A.~Tilquin, \emph{Cosmological parameter extraction and biases from type ia supernova magnitude evolution}, \href{https://doi.org/10.1051/0004-6361/200912811}{\emph{A\&A} {\bfseries 506} (2009) 1095}.

\bibitem{Tutusaus_2017}
I.~Tutusaus, B.~Lamine, A.~Dupays and A.~Blanchard, \emph{Is cosmic acceleration proven by local cosmological probes?}, \href{https://doi.org/10.1051/0004-6361/201630289}{\emph{A\&A} {\bfseries 602} (2017) A73}.

\bibitem{Tutusaus_2018}
I.~Tutusaus, B.~Lamine and A.~Blanchard, \emph{Model-independent cosmic acceleration and redshift-dependent intrinsic luminosity in type-ia supernovae}, \href{https://doi.org/10.1051/0004-6361/201833032}{\emph{A\&A} {\bfseries 625} (2019) A15}.

\bibitem{Yijung_2020}
Y.~Kang, Y.-W.~Lee, Y.-L.~Kim, C.~Chung and C.H.~Ree, \emph{Early-type host galaxies of type ia supernovae. {II}. evidence for luminosity evolution in supernova cosmology}, \href{https://doi.org/10.3847/1538-4357/ab5afc}{\emph{The Astrophysical Journal} {\bfseries 889} (2020) 8}.

\bibitem{Benisty_2022}
D.~Benisty, J.~Mifsud, J.L.~Said and D.~Staicova, \emph{On the robustness of the constancy of the supernova absolute magnitude: Non-parametric reconstruction \& bayesian approaches},  2022.

\bibitem{Riess_2018}
A.G.~Riess, S.A.~Rodney, D.M.~Scolnic, D.L.~Shafer, L.-G.~Strolger, H.C.~Ferguson et~al., \emph{Type ia supernova distances at redshift $>$1.5 from the hubble space telescope multi-cycle treasury programs: The early expansion rate}, \href{https://doi.org/10.3847/1538-4357/aaa5a9}{\emph{The Astrophysical Journal} {\bfseries 853} (2018) 126}.

\bibitem{Pantheon}
D.M.~Scolnic, D.O.~Jones, A.~Rest, Y.C.~Pan, R.~Chornock, R.J.~Foley et~al., \emph{The complete light-curve sample of spectroscopically confirmed {SNe} ia from pan-{STARRS}1 and cosmological constraints from the combined pantheon sample}, \href{https://doi.org/10.3847/1538-4357/aab9bb}{\emph{The Astrophysical Journal} {\bfseries 859} (2018) 101}.

\bibitem{Howlett_2017}
C.~Howlett, A.S.G.~Robotham, C.D.P.~Lagos and A.G.~Kim, \emph{Measuring the growth rate of structure with type {IA} supernovae from {LSST}}, \href{https://doi.org/10.3847/1538-4357/aa88c8}{\emph{The Astrophysical Journal} {\bfseries 847} (2017) 128}.

\bibitem{Brout_2022}
D.~Brout, D.~Scolnic, B.~Popovic, A.G.~Riess, A.~Carr, J.~Zuntz et~al., \emph{The pantheon$+$ analysis: Cosmological constraints}, \href{https://doi.org/10.3847/1538-4357/ac8e04}{\emph{The Astrophysical Journal} {\bfseries 938} (2022) 110}.

\end{thebibliography}\endgroup

\end{document}